\documentclass[12pt]{article}
\usepackage{amsmath}
\usepackage{cite}
\usepackage[dvipdfmx]{graphicx}
\usepackage{color}
\begin{document}
\begin{flushright}
\end{flushright}
\vspace*{1cm}

\renewcommand\thefootnote{\fnsymbol{footnote}}
\begin{center} 
 {\Large\bf Inflation caused by a potential valley with $CP$ violation}
\vspace*{1cm}

{\Large Daijiro Suematsu}\footnote{emeritus professor, 
e-mail:~suematsu@hep.s.kanazawa-u.ac.jp}
\vspace*{0.5cm}\\

{\it Institute for Theoretical Physics, Kanazawa University, 
Kanazawa 920-1192, Japan}

\end{center}
\vspace*{1.5cm} 

\noindent
{\Large\bf Abstract}\\
We propose an inflation scenario caused by an inflaton identified with a potential
 valley which brings about $CP$ violation in the standard model (SM). 
If singlet scalars have nonminimal couplings with the Ricci scalar in a suitable way, 
favorable inflation could be derived through the potential 
valley composed of such scalars.
We study dynamics of the scalars during and after the inflation, which suggests
that it could be described as a single field inflation approximately.  
If these scalars make suitable ingredients of the SM couple with 
vectorlike fermions and
right-handed neutrinos, a CKM phase could be induced and also leptogenesis 
could occur successfully even at a rather low reheating temperature. 
 
\newpage
\setcounter{footnote}{0}
\renewcommand\thefootnote{\alph{footnote}}

\section{Introduction}
Existence of exponential space expansion at an early stage of the universe
called inflation is now believed seriously based on observational data 
of the CMB fluctuation \cite{cmb1,cmb2,cmb3,planck18}.
Inflation is usually considered to be induced by some slowly rolling scalar 
field called inflaton \cite{inf1,inf2,inf3}. Clarification of its candidate is a crucial 
problem in cosmology and astroparticle physics. Moreover, it is also an unavoidable problem 
from a viewpoint of the extension of the SM. 
If Higgs scalar has a nonminimal coupling with the Ricci scalar 
curvature \cite{nonminimal1,nonminimal3,nonminimal4,nonminimal2}, it could be a promising 
candidate of inflaton in the SM \cite{higgsinf1,higgsinf3,higgsinf4,higgsinf2}.
Such a possibility has been studied extensively and several discussions on its 
applicability have been done \cite{stab1,stab2,stab3,stab4,stab5,unitarity1,
unitarity2,unitarity3,unitarity4,
unitarity5,qeffect1,qeffect2,qeffect3,spike}.  Under this situation, 
it seems to be an interesting subject 
to find out an alternative candidate 
of the inflaton in models extended to explain unsolved problems in the SM. 
In this sense, it may be instructive to consider an inflaton candidate 
from a viewpoint of the origin of $CP$ violation in the SM \cite{cpinf}.

Usually, $CP$ symmetry in the SM is considered to be explicitly violated 
through complex Yukawa couplings which fix mass matrices ${\cal M}_u$ 
and ${\cal M}_d$ of up-type and down-type quarks \cite{km}.
In this case, a $\theta$ parameter in the QCD sector \cite{theta1,theta2} is irrelevant to
the complex phases of the Yukawa couplings. As a result, we cannot understand 
why $\bar\theta=\theta+{\rm arg}({\det({\cal M}_u{\cal M}_d)})$ takes an extremely small value
less than $10^{-10}$, which is required by the experimental bound of the neutron electric dipole
moment \cite{nedm}. This is known as the strong $CP$ problem in the SM \cite{strongcp1,strongcp2}.
It can be solved by introducing the Peccei-Quinn symmetry \cite{pq1,pq2}. Its spontaneous 
breaking can solve the problem by making $\bar\theta$ a dynamical variable called 
axion \cite{axion1,axion2,axion3,axion4,axion5,axion6}.  
However, if the $CP$ is considered as an exact symmetry of the model 
originally but it is spontaneously broken at some scale, an alternative solution for the
strong $CP$ problem is known to be constructed. In that case,
the $CP$ symmetry imposes $\theta=0$. Moreover, if its spontaneous breaking causes complex 
phases of the Yukawa couplings satisfying ${\rm arg}({\det({\cal M}_u{\cal M}_d)})=0$, the strong $CP$
problem can be solved \cite{nb1,nb2,nb3}. 
It makes a model based on the spontaneous $CP$ violation worthy to study.
Introduction of singlet scalars makes such a scenario possible \cite{spcp}. 
If they couple with the Ricci scalar nonminimally, they could cause 
slow-roll inflation successfully just as the Higgs inflation \cite{sinf1,sinf2,
ext-s1,ext-s2,ext-s3,ext-s4,ext-s5,ext-s6,hs}. 
One of important differences from the Higgs inflation is that their nonminimal 
couplings can take a value of $O(1)$ consistently with the CMB data.
In this paper, we study the possibility of such singlet scalars as an inflaton 
candidate, which could be relevant to the origin of the $CP$ violation and several 
phenomenological issues in the SM.

Remaining parts of the paper are organized as follows.
In section 2, we describe a scalar sector which causes 
the spontaneous $CP$ violation. We fix an inflaton candidate in a space of 
scalar fields and discuss the inflation brought about by it. 
We also discuss reheating after the inflation.
In section 3 we describe characteristic low energy phenomena brought 
about in this inflation scenario, that is, the generation of $CP$ phases in the
CKM and PMNS matrices, and a possibility of low scale leptogenesis.
The paper is summarized in section 4.

\section{Inflation caused by singlet scalars}
\subsection{Spontaneous $CP$ violation}
We consider a $CP$ invariant model with two singlet scalars that are  
a real singlet scalar $\sigma$ and a complex scalar $S$ \cite{ps1,ps2}.
If we impose discrete symmetry $Z_4\times Z_4^\prime$ whose charges 
are assigned these scalars as $\sigma~(2,2)$\footnote{Since $\sigma$ is the real field, 
only the parity type transformation is allowed. We note that this charge realizes it. } 
and  $S~(2,0)$,
the model is characterized by the following scalar potential invariant under 
the imposed symmetry:
\begin{eqnarray}
V_0(S,S^\dagger,\sigma)&=&\frac{1}{4}\kappa_\sigma\sigma^4
+\kappa_S(S^\dagger S)^2
+\frac{1}{2}\kappa_{\sigma S}\sigma^2(S^\dagger S) 
+\frac{1}{2}m_\sigma^2\sigma^2+m_S^2S^\dagger S \nonumber \\
&+&\alpha(S^4+S^{\dagger 4})+\frac{1}{2}\beta\sigma^2(S^2+S^{\dagger 2}).
\label{model}
\end{eqnarray} 
The SM contents are assumed to have no charge of this symmetry. 
Since $CP$ invariance is assumed in the model, all parameters 
in the Lagrangian are real. 
The second line in eq.~(\ref{model}) is composed of terms which violate 
the $S$ number.
For a while, we focus only on $\sigma$ and $S$ parts and neglect their 
couplings with the ordinary Higgs scalar.

This potential can be rewritten in a form such as
\begin{eqnarray}
V_0(\tilde S,\sigma,\rho)&=&\frac{\tilde\kappa_\sigma}{4}(\sigma^2-w^2)^2+
\frac{\tilde\kappa_S}{4}(\tilde S^2-u^2)^2+
\frac{\kappa_{\sigma S}}{4}(\sigma^2-w^2)(\tilde S^2-u^2) \nonumber\\
&+&\alpha(\tilde S^2\cos 2\rho+\frac{\beta}{4\alpha}\sigma^2)^2,
\label{potv}
\end{eqnarray}
where $\tilde S$ and $\rho$ are defined by 
$S=\frac{\tilde S}{\sqrt 2}e^{i\rho}$. $w$ and $u$ are
vacuum expectation values (VEVs) of $\sigma$ and $\tilde S$, and
a constant term in the potential is assumed to be zero. 
Coupling constants $\tilde\kappa_\sigma$ 
and $\tilde \kappa_S$ are defined as
\begin{equation}
\tilde\kappa_\sigma=\kappa_\sigma-\frac{\beta^2}{4\alpha}, \qquad
\tilde\kappa_S=\kappa_S-2\alpha.
\label{tilcoup}
\end{equation}
Stability of the potential (\ref{potv}) is guaranteed under the conditions
\begin{equation}
\tilde\kappa_\sigma,~ \tilde\kappa_S>0, \qquad  
4\tilde\kappa_\sigma\tilde\kappa_S>\kappa_{\sigma S}^2.
\label{stab1}
\end{equation} 
If the VEVs $w$ and $u$ satisfy 
$\frac{w^2}{u^2}\left|\frac{\beta}{4\alpha}\right|\le 1$, 
the minimum of $V_0(\tilde S,\sigma,\rho)$ is realized for $\rho$ such that 
$\cos 2\rho=-\frac{\beta}{4\alpha}\frac{w^2}{u^2}$ and
the $CP$ symmetry could be spontaneously broken.
Moreover, if $w$ and $u$ are supposed to be larger than the weak scale,
desirable $CP$ violation could be caused in the SM sector from this origin
by introducing suitable mediators and interactions additionally. 
The lower bound of these VEVs could be given by a unitarity constraint
on the CKM matrix and also  a requirement for successful leptogenesis.  
We discuss this point in section 3.

\subsection{Inflaton defined as a $CP$ violating potential valley}
In a model where a scalar field couples non-minimally with the Ricci scalar, 
inflation of the universe is known to be caused by that scalar \cite{nonminimal1,nonminimal3,nonminimal4,nonminimal2}.
If the singlet scalars $S$ and $\sigma$ in the present model
couple with the Ricci scalar, they are expected to play a role of inflaton. 
Action relevant to the inflation is given in the Jordan frame as \cite{cpinf}
\begin{eqnarray}
S_J &=& \int d^4x\sqrt{-g} \left[-\frac{1}{2}M_{\rm pl}^2R -
\frac{1}{2}\xi_\sigma\sigma^2 R-
\xi_{S_1} S^\dagger S R -\frac{\xi_{S_2}}{2}(S^2+S^{\dagger 2})R\right. \nonumber \\
&+&\left.\frac{1}{2}\partial^\mu \sigma\partial_\mu \sigma
+\partial^\mu S^\dagger \partial_\mu S - V_0(S, S^\dagger, \sigma) \right],
\label{inflag}
\end{eqnarray}
where $M_{\rm pl}$ is the reduced Planck mass.
Nonminimal couplings can be rewritten as
\begin{equation}
\frac{1}{2}\left[\xi_\sigma\sigma^2+(\xi_{S_1}+\xi_{S_2})S_R^2+(\xi_{S_1}-\xi_{S_2})S_I^2\right]R,
\end{equation}
where $S_R$ and $S_I$ are real and imaginary parts of $S$ and $S=\frac{1}{\sqrt 2}(S_R+iS_I)$.
We focus the present study on the cases where only one real scalar $\chi$ is allowed
to have the nonminimal coupling other than $\sigma$.
If we express this coupling as $\frac{1}{2}\xi_\chi\chi^2R$ 
by assuming a certain condition for $\xi_{S_1}$ and $\xi_{S_2}$, 
they can be described as\footnote{We can consider another possibility such that
$\chi=S_R$ in the case $\xi_{S_1}=\xi_{S_2}$. 
However, we do not discuss it here since the inflation feature is similar to the case (ii).  }  
\begin{equation}
{\rm (i)}~\xi_{S_2}=0;\quad  \chi=\tilde S,\quad \xi_\chi=\xi_{S_1}, \qquad
{\rm (ii)}~\xi_{S_1}=-\xi_{S_2};\quad  \chi=S_I,\quad  \xi_\chi=\xi_{S_1}-\xi_{S_2}.
\label{defx}
\end{equation}
We note that the $S$ number of the nonminimal coupling 
is conserved in the case (i) but not in the case (ii).
Although these give the same low energy effective theory described 
by eq.~(\ref{potv}), a difference could appear in inflation phenomena. 
In the following parts, we confine our study to the case $\xi_\chi,~\xi_\sigma>0$.

We consider the conformal transformation for a metric tensor in the Jordan 
frame \cite{nonminimal3,nonminimal4}
\begin{equation}
\tilde g_{\mu\nu}=\Omega^2g_{\mu\nu}, \qquad
\Omega^2=1+\frac{(\xi_\sigma\sigma^2+\xi_\chi \chi^2)}{M_{\rm pl}^2}.
\end{equation}
After this transformation to the Einstein frame where the Ricci scalar term takes a canonical form,
the action can be written as \cite{sinf1,sinf2}
\begin{eqnarray}
S_E &=& \int d^4x\sqrt{-\tilde g} \Big[-\frac{1}{2}M_{\rm pl}^2\tilde R
+\sum_{i=R,I}\left(\frac{f_i}{2} \partial^\mu S_i \partial_\mu S_i +
 f_{\sigma i}\partial^\mu\sigma\partial_\mu S_i\right) \nonumber \\
   &+&\frac{1}{2\Omega^4}\left(\Omega^2+6\xi_\sigma^2
\frac{\sigma^2}{M_{\rm pl}^2}\right)\partial^\mu\sigma \partial_\mu \sigma
+f_{RI}\partial^\mu S_R \partial_\mu S_I
  -V \Big],
\end{eqnarray}
where the potential $V$ is given by 
$V=\frac{1}{\Omega^4}V_0(S, S^\dagger, \sigma)$.
Since $\Omega^2$ has a different dependence on $S_R$ and $S_I$ in each case, 
the kinetic terms of $S_{R,I}$ have different expressions 
\small
\begin{eqnarray}
&&{\rm (i)}~f_i=\frac{1}{\Omega^4}\left(\Omega^2+6\xi_\chi^2 \frac{S_i^2}{M_{\rm pl}^2}\right), 
\quad  f_{RI}=\frac{1}{\Omega^4}\frac{6\xi_\chi^2S_RS_I}{M_{\rm pl}^2}, \quad
f_{\sigma i}=\frac{1}{\Omega^4}\frac{6\xi_\sigma\xi_\chi\sigma S_i}{M_{\rm pl}^2},
 \nonumber \\
&&{\rm (ii)}~f_R=\frac{1}{\Omega^2}, \quad 
f_I=\frac{1}{\Omega^4}\left(\Omega^2+6\xi_\chi^2 \frac{S_I^2}{M_{\rm pl}^2}\right), 
 \quad 
f_{\sigma I}=\frac{1}{\Omega^4}\frac{6\xi_\sigma\xi_\chi\sigma S_I}{M_{\rm pl}^2},
\quad f_{RI}=f_{\sigma R}=0 ,
\label{infkin}
\end{eqnarray}
\normalsize
where $i=R,I$ should be understood.
We neglect $w$ and $u$ in $V_0(S, S^\dagger, \sigma)$ for a while 
since they are much smaller than $O(M_{\rm pl})$ 
that is supposed to be a value of $\sigma$ and $\chi$ during the inflation. 
If we express $\sigma$ and $\chi$ as $\sigma=\tilde\chi\cos\varphi$ and 
$\chi=\tilde\chi\sin\varphi$, 
the potential $V$ in the Einstein frame can be rewritten as
\begin{equation}
V=\frac{M_{\rm pl}^4}{4}
\frac{\tilde\kappa_S\sin^4\varphi+
\tilde\kappa_\sigma\cos^4\varphi
+\kappa_{\sigma S}\sin^2\varphi\cos^2\varphi +\tilde V(\varphi,\rho)} 
{(\xi_\sigma\cos^2\varphi+\xi_\chi\sin^2\varphi)^2} 
\label{vk}
\end{equation}
at large field regions satisfying $\xi_\sigma\sigma^2+\xi_\chi\chi^2> M_{\rm pl}^2$.
 $\tilde V$ is expressed in each case as
\begin{eqnarray}
{\rm (i)}~~\tilde V(\varphi,\rho)&=&\alpha\sin^2\varphi\left(\cos 2\rho+\frac{\beta}{4\alpha}\cot^2\varphi\right)^2, \nonumber \\ 
{\rm (ii)}~~\tilde V(\varphi,\rho)&=& \tilde\kappa_S\sin^4\varphi\cot^4\rho
+2\tilde\kappa_S\sin^2\varphi\cot^2\rho+\tilde\kappa_{\sigma S}\sin^2\varphi\cos^2\varphi\cot^2\rho \nonumber \\
&+&\alpha\sin^2\varphi\left(\cot^2\rho-1+
\frac{\beta}{4\alpha}\cot^2\varphi\right)^2.
\end{eqnarray} 
$V$ is found to be independent of $\tilde\chi$ and take a constant value at the potential 
minimum as long as $\varphi$ and $\rho$ are constant there.  
It suggests that $\tilde\chi$ could play a role of the slow roll inflaton.

Minima of this potential are found from the condition
$\frac{\partial V}{\partial \rho}=\frac{\partial V}{\partial \varphi}=0$ and they 
form valleys in a space of the fields $S_R,~S_I$ and $\sigma$.
In the case (i),  the minimum for $\rho$ can be realized at 
$\cos 2\rho=-\frac{\beta}{4\alpha}\cot^2\varphi$ for suitable parameters, 
and $V$ has three types of valley in the $\varphi$ direction \cite{cpinf}. 
One of these valleys, which is studied in this paper,\footnote{
Other valleys are given for $\varphi=0$ and $\frac{\pi}{2}$.
Inflation in the case $\varphi=\frac{\pi}{2}$ has been discussed  in
the different context \cite{ps1,ps2}.}
is realized at 
\begin{equation}
\sin^2\varphi=\frac{2\tilde\kappa_\sigma\xi_\chi-\kappa_{\sigma S}\xi_\sigma}
{(2\tilde\kappa_S\xi_\sigma-\kappa_{\sigma S}\xi_\chi)
+(2\tilde\kappa_\sigma\xi_\chi-\kappa_{\sigma S}\xi_\sigma)}
\label{varphi}
\end{equation}
under the condition
\begin{equation}
 2\tilde\kappa_\sigma\xi_\chi>\kappa_{\sigma S}\xi_\sigma, \qquad 
2\tilde\kappa_S\xi_\sigma>\kappa_{\sigma S}\xi_\chi.
\label{cond}
\end{equation} 
This condition is automatically satisfied 
for any positive $\xi_\chi$ and $\xi_\sigma$ 
under the condition (\ref{stab1}) with $\kappa_{\sigma S}<0$.
In such a case, both $\varphi$ and $\rho$ are kept to be constant in this valley.
Although the kinetic term mixing in eq.~(\ref{inflag}) cannot be neglected 
generally, it can be safely neglected in a case $\xi_\chi\gg \xi_\sigma$. 
We confine our study to such a case here. 
If we additionally assume that the relevant couplings satisfy
\begin{equation}
\kappa_{\sigma S}<0, \qquad \tilde\kappa_S\ll |\kappa_{\sigma S}|\ll \tilde\kappa_\sigma,
\label{ks}
\end{equation}
$\sin\varphi$ can be expressed as 
$\sin^2\varphi\simeq 1+\frac{\kappa_{\sigma S}}{2\tilde\kappa_\sigma}$.
Nature of the inflaton $\chi$ is fixed by the parameters $\tilde\kappa_S$, 
$\tilde\kappa_\sigma$ and $\kappa_{\sigma S}$.
$S_R,~S_I$ and $\sigma$ satisfy
\begin{eqnarray}
S_R^2=\frac{1+\left(1+\frac{\beta}{4\alpha}\right)\frac{\kappa_{\sigma S}}{2\tilde\kappa_\sigma}}
{1+\left(1-\frac{\beta}{4\alpha}\right)\frac{\kappa_{\sigma S}}{2\tilde\kappa_\sigma}}S_I^2, \qquad
\sigma^2=\frac{-\frac{\kappa_{\sigma S}}{\tilde\kappa_\sigma}}{1+\left(1-\frac{\beta}{4\alpha}\right)\frac{\kappa_{\sigma S}}{2\tilde\kappa_\sigma}}S_{I}^2
\label{valley}
\end{eqnarray}
at the bottom of this valley. 

In the case (ii), $\cot\rho\not=0$ requires a rather restricted 
value for $\xi_\chi/\xi_\sigma$ and the formula for $\varphi$ 
at the potential minimum is complicated. 
However, in a case $\cot\rho=0$, 
we can find a simplified expression such that
\begin{equation}
\sin^2\varphi=\frac{(2\tilde\kappa_S+\frac{\beta^2}{2\alpha})\xi_\chi
-(\kappa_{\sigma S}-2\beta)\xi_\sigma}
{(2\tilde\kappa_\sigma-\kappa_{\sigma S}+
\frac{\beta^2}{2\alpha}-2\beta)\xi_\chi+
(2\tilde\kappa_S-\kappa_{\sigma S}+8\alpha+2\beta)\xi_\sigma}.
\label{varphi2}
\end{equation}
If $\xi_\chi\gg\xi_\sigma$ and $\alpha\simeq\beta$ are assumed to be satisfied together with
eq.~(\ref{ks}),
this gives $\sin^2\varphi\simeq 1+\frac{\kappa_{\sigma S}}{2\tilde\kappa_\sigma}$
that is the same as one in the case (i).
Even in the case that a minimum is realized at $\cot\rho\not=0$,
only $\cot\rho\simeq 0$ is allowed under the present assumption 
and $\sin^2\varphi\simeq 1$ is fulfilled. 
Therefore, we adopt the case $\cot\rho=0$ as a typical example in the case (ii).
In this case, $S_R,~S_I$ and $\sigma$ satisfy 
\begin{equation}
S_R=0, \qquad \sigma^2=\frac{-(\kappa_{\sigma S}+2\beta)}
{2\tilde\kappa_\sigma+\frac{\beta^2}{2\alpha}}S_I^2
\label{valley2}
\end{equation}
at the bottom of the valley.

If we rewrite the kinetic term in eq.~(\ref{inflag}) taking account of  
$\xi_\chi\gg\xi_\sigma$, 
a canonically normalized inflaton $\hat\chi$ is found to satisfy 
\begin{equation}
\Omega^2\frac{d\hat\chi}{d\chi}=\sqrt{\gamma\Omega^2+6\xi_\chi^2
\frac{\chi^2}{M_{\rm pl}^2}},  
\label{tchi}
\end{equation}
where $\gamma$ is defined as
\begin{equation}
\gamma\equiv\frac{1}{\sin^2\varphi}\simeq 1-
\frac{\kappa_{\sigma S}}{2\tilde\kappa_\sigma}.
\end{equation}
If we use $\gamma\simeq 1$ that is satisfied along this valley, 
$\hat\chi$ can be derived as a solution of eq.~(\ref{tchi}).  It is found to be given as
\begin{equation}
\frac{\hat\chi}{M_{\rm pl}}=-\sqrt 6~{\rm arcsinh}\left(\frac{\sqrt\frac{6}{\gamma}
\frac{\xi_\chi\chi}{M_{\rm pl}}}
{\sqrt{1+\frac{\xi_\chi}{M_{\rm pl}^2}\chi^2}}\right)+
\sqrt\frac{\gamma+6\xi_\chi}{\xi_\chi}{\rm arcsinh}
\left(\frac{\sqrt{\xi_\chi(1+\frac{6}{\gamma}\xi_\chi)}\chi}{M_{\rm pl}}\right).
\label{etchi}
\end{equation}
The potential of $\hat\chi$ can be fixed through 
$V(\hat\chi)=\frac{1}{\Omega^4}V(\sigma,\chi)$ by using this relation.
It can be approximately expressed as 
\begin{eqnarray}
V(\hat\chi)=\left\{ \begin{array}{ll}
\frac{\hat\kappa_S}{4\xi_\chi^2}M_{\rm pl}^4 & \quad M_{\rm pl}<\hat\chi  \\
\frac{\hat\kappa_S}{6\xi_\chi^2}M_{\rm pl}^2\hat\chi^2  
&\quad \frac{M_{\rm pl}}{\xi_\chi}<\hat\chi< M_{\rm pl} \\
 \frac{\hat\kappa_S}{4}\hat\chi^4 &\quad \hat\chi < \frac{M_{\rm pl}}{\xi_\chi},
\end{array}\right.
\label{oscilpot}
\end{eqnarray}
where $\hat\kappa_S\equiv\tilde\kappa_S-\frac{\kappa_{\sigma S}^2}
{4\tilde\kappa_\sigma}$ and $\hat\kappa_S>0$ is guaranteed 
by eq.~(\ref{stab1}). 

Here we describe constraints on the model parameters from the
CMB observation.
Slow-roll parameters of the inflation can be estimated by using 
eq.~(\ref{tchi}) as \cite{inf1,inf2}
\begin{equation}
\epsilon\equiv\frac{M_{\rm pl}^2}{2}\left(\frac{V^\prime}{V}\right)^2
=\frac{8M_{\rm pl}^4}{\gamma\xi_\chi\left(1+\frac{6}{\gamma}\xi_\chi\right)\chi^4}, \qquad
\eta\equiv M_{\rm pl}^2\frac{V^{\prime\prime}}{V}=
-\frac{8M_{\rm pl}^2}{\gamma\left(1+\frac{6}{\gamma}
\xi_\chi\right)\chi^2},
\end{equation}
where $V^\prime$ stands for $\frac{dV}{d\hat\chi}$.
The $e$-foldings number ${\cal N}_k$ from a time when a scale 
$k$ exits the horizon to the end of inflation is expressed by applying eq.~(\ref{tchi}) as 
\begin{eqnarray}
{\cal N}_k=\frac{1}{M_{\rm pl}^2}\int^{\hat\chi_k}_{\hat\chi_{\rm end}}\frac{V}{V^\prime}d\hat\chi
=\frac{1}{8M_{\rm pl}^2}(\gamma+6\xi_\chi)(\hat\chi_k^2-\hat\chi_{\rm end}^2)
-\frac{3}{4}\ln\frac{M_{\rm pl}^2+\xi_\chi \hat\chi_k^2}
{M_{\rm pl}^2+\xi_\chi \hat\chi_{\rm end}^2},
\label{efold}
\end{eqnarray}
where $\hat\chi_{\rm end}$ is an inflaton value at the end of 
inflation that is estimated
from $\epsilon\simeq 1$.
If we use this ${\cal N}_k$, the slow-roll parameters of the model 
 are found to be approximated as 
\begin{equation}
\epsilon\simeq \frac{3}{4{\cal N}_k^2}, \qquad \eta\simeq -\frac{1}{{\cal N}_k}.
\label{slowroll}
\end{equation}
These predict favorable values for the scalar spectral index $n_s$ and the 
tensor-to-scalar ratio $r$ for ${\cal N}_k=50-60$. 
We return this point at the last part of section 2.3.

On the other hand, by using eqs.~(\ref{etchi}) and (\ref{efold}), 
the field value of inflaton during the inflation can be  expressed as 
$\hat\chi_k=\frac{\sqrt 6}{2}M_{\rm pl}\ln(32\xi_\chi{\cal N}_k)$ 
and the potential $V_k(\equiv V(\hat\chi_k))$ takes 
a constant value as shown in eq.~(\ref{oscilpot}).
If we use $\epsilon=1$ at the end of inflation, the inflaton potential is 
estimated as $V_{\rm end}(\equiv V(\hat\chi_{\rm end}))\simeq 0.072
\frac{\hat\kappa_S}{\xi_\chi^2}M_{\rm pl}^4$. 
Spectrum of the CMB density perturbation predicted by the slow roll inflation 
is known to be expressed as \cite{inf1,inf2}
\begin{equation}
{\cal P}(k)=A_s\left(\frac{k}{k_\ast}\right)^{n_s-1},  \qquad
A_s=\frac{V}{24\pi^2M_{\rm pl}^4\epsilon}\Big|_{k_\ast}. 
\label{power}
\end{equation}
If we use the Planck data $A_s=(2.101^{+0.031}_{-0.034})\times 10^{-9}$ 
at $k_\ast=0.05~{\rm Mpc}^{-1}$ \cite{planck18}, we find the relation 
\begin{equation}
\hat\kappa_S\simeq 4.13\times 10^{-10}\xi_\chi^2 
\left(\frac{60}{{\cal N}_{k_\ast}}\right)^2,
\label{kap}
\end{equation}
and the Hubble parameter satisfies $H_I=1.4\times 10^{13}
\left(\frac{60}{{\cal N}_{k_\ast}}\right)$~GeV during the inflation. 
Since the inflaton stays at the bottom of the valley at this period, 
field values of $S_R, S_I$ and $\sigma$ at the end of inflation 
can be fixed by eq.~(\ref{valley}) or (\ref{valley2}) in each case.   

We note that the relevant scalars need not to be at the bottom of the valley
initially. One of orthogonal components to $\hat\chi$ has mass 
$m_{\hat\chi_\perp}^2\simeq \frac{|\kappa_{\sigma S}|M_{\rm pl}^2}
{2\xi_\chi^2}$ during the inflation. 
Since the Hubble parameter $H_I$ is expressed as 
$H_I^2=\frac{\hat\kappa_S M_{\rm pl}^2}{12\xi_\chi^2}$ at the same period,
$H_I < m_{\hat\chi_\perp}$ is satisfied under the condition (\ref{ks}). 
Mass of another orthogonal scalar is estimated as 
$m^2_\rho\simeq \frac{8\alpha M_{\rm pl}^2}{\xi_\chi}$ 
for $\cos 2\rho\sim 0$ in the case (i), and 
$m_{S_R}^2\simeq\frac{\hat\kappa_S M_{\rm pl}^2}{\xi_\chi}$ in the case (ii).
As long as $\xi_\chi>1$ is satisfied, both of these could be larger than $H_I^2$ 
for suitable coupling values without causing any other problem.
These suggest that they reach the bottom of the valley and begin to roll 
along it within a few Hubble time even if the relevant scalars start rolling 
at points displaced from the bottom of the valley.
This fact guarantees the isocurvature perturbation caused by other components
than $\hat\chi$ to be negligibly small. 

\begin{figure}[t]
\begin{center}
\includegraphics[width=7.5cm]{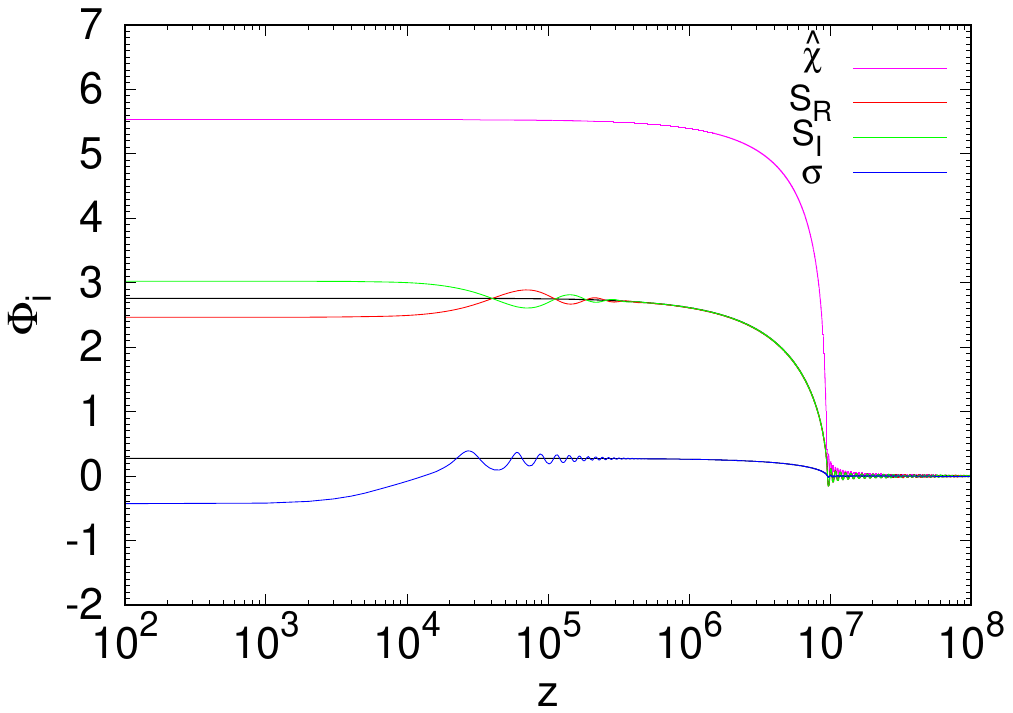}
\hspace*{5mm}
\includegraphics[width=7.5cm]{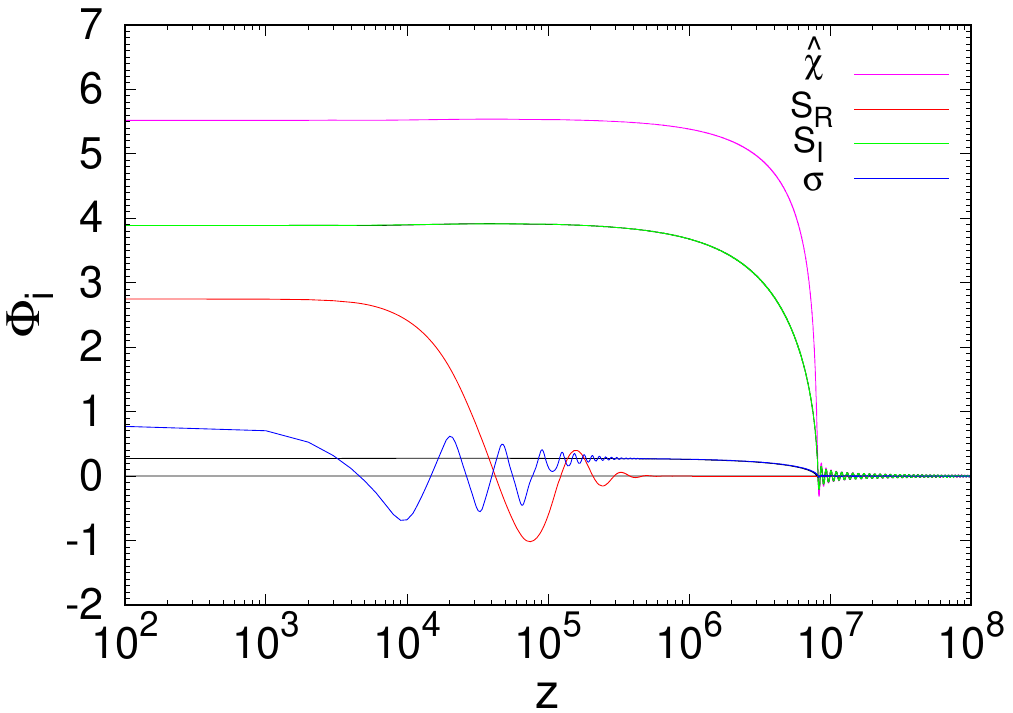}\\
\includegraphics[width=7.5cm]{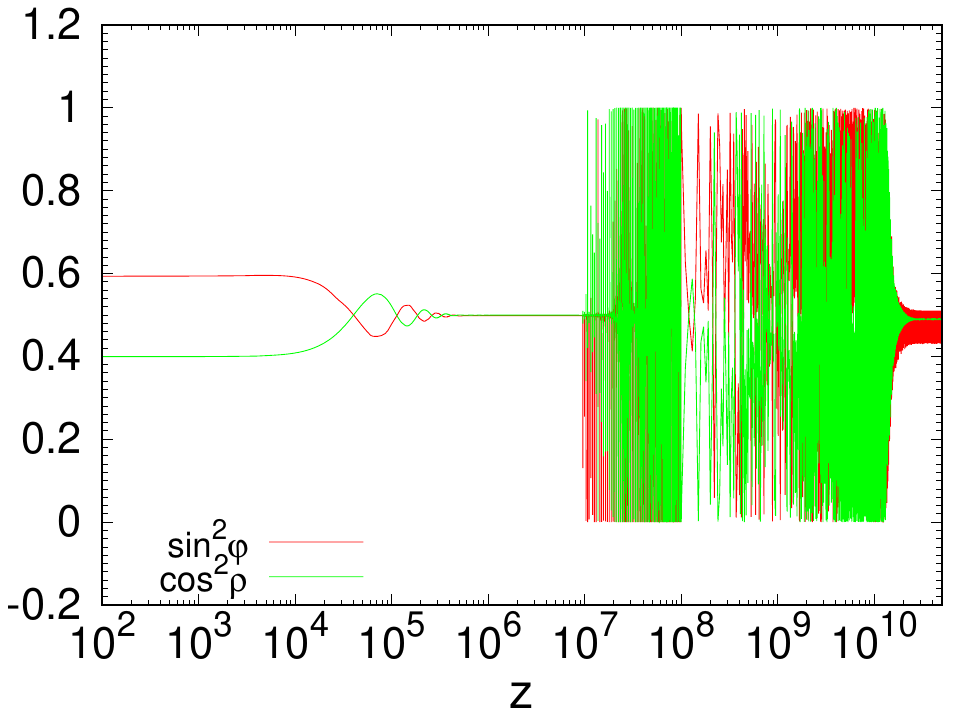}
\hspace*{5mm}
\includegraphics[width=7.5cm]{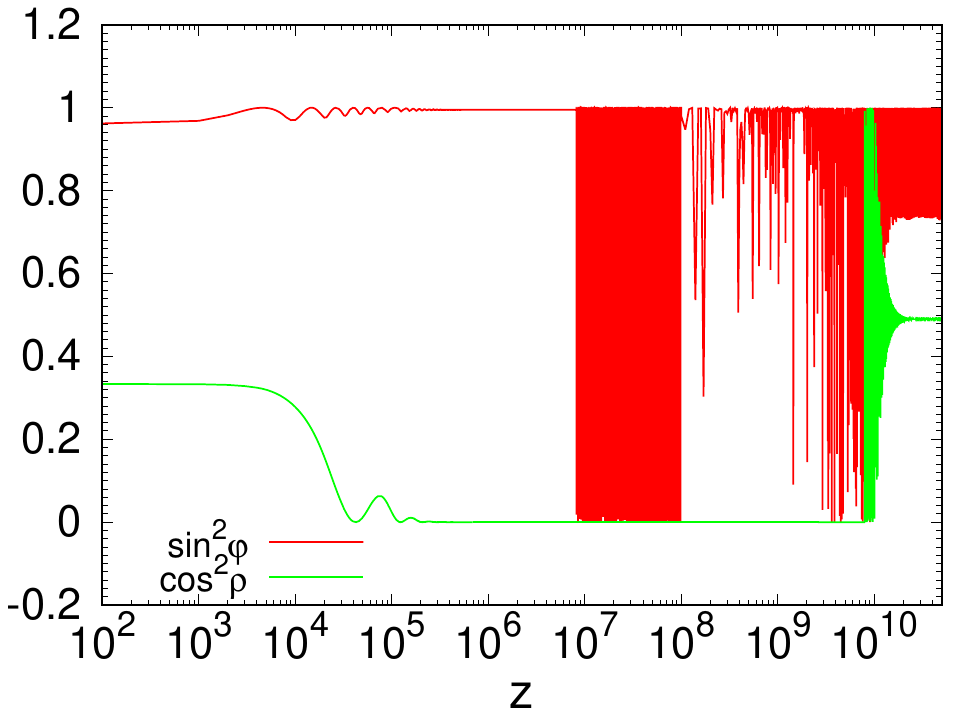}\\
\end{center}
\vspace*{-2mm}
\footnotesize{{\bf Fig.~1}~~Upper panels: Time evolution of scalar fields 
$S_R$, $S_I, \sigma$ and $\hat\chi$ as functions of a dimensionless time 
$z=M_{\rm pl}t$. Each field is described by a Planck mass unit.
Initial values are taken to be deviated from the bottom of the valley
described by eqs.~(\ref{valley}) and (\ref{valley2}).
As references, using the black lines we plot solutions for the initial 
values given by these equations.
The left and right panels are the ones for the case (i) and (ii), respectively.
Lower panels: Time evolution of $\sin^2\varphi$ and $\cos^2\rho$ obtained for 
the shifted initial values used in the upper panels. 
}
\end{figure}

We can examine the validity of this single field treatment  
by solving coupled field equations for $S_R, S_I$ and $\sigma$ numerically.
They are given as
\begin{equation}
\frac{d^2\Phi_i}{dt^2}+3H\frac{d\Phi_i}{dt}+\frac{dV}{d\Phi_i}=0,
\label{ceofm}
\end{equation}
where $\Phi_i$ stands for $S_R, S_I$ and $\sigma$. The potential $V$ is given 
as $V=\frac{1}{\Omega^4}V_0$ by rewriting eq.~(\ref{potv}) as
\begin{eqnarray}
V_0(S_R,S_I,\sigma)&=&\frac{\hat\kappa_S}{4}(S_R^2+S_I^2-u^2)^2
+\frac{\tilde\kappa_\sigma}{4}\left\{\sigma^2-w^2+\frac{\kappa_{\sigma S}}
{2\tilde\kappa_\sigma}(S_R^2+S_I^2-u^2)\right\}^2  \nonumber \\
&+&\alpha\left(S_R^2-S_I^2+\frac{\beta}{4\alpha}\sigma^2\right)^2.
\label{lowp}
\end{eqnarray}
Since $\hat\kappa_S$ has to satisfy eq.~(\ref{kap}),
it is fixed to $\hat\kappa_S=1.23\times 10^{-8} $ for 
$\xi_\chi=5$ and ${\cal N}_k=55$.
As an example, we fix other parameters 
to\footnote{We find that $\kappa_S$ is fixed at $\kappa_S=1.02\times 10^{-8}$ by these. }
\begin{equation}
\xi_\sigma=10^{-3}, \quad \kappa_\sigma=10^{-4.5}, \quad 
\kappa_{\sigma S}~=-10^{-6.5}, \quad
\alpha=\beta=10^{-1}\hat\kappa_S.
\label{para}
\end{equation}
These values are confirmed to fulfill the imposed conditions 
(\ref{stab1}), (\ref{ks}) and (\ref{kap}).  
Using these parameters, we solve eq. (\ref{ceofm}) to find the time evolution 
of $S_R,~S_I$ and $\sigma$.
Their initial values are fixed so as to deviate from the potential minimum
taking account of eqs.~(\ref{valley}) and (\ref{valley2}). 
The results are plotted in Fig.~1. 
Figures show that each field converges to the values which give
constant $\varphi$ and $\rho$ at $z\simeq 7.7\times 10^5$ in the case (i) and
$6.1\times 10^5$ in the case (ii), which 
are sufficiently earlier than the time when ${\cal N}_k=55$ is realized 
at the end of inflation $z_{\rm end}=9.6\times 10^6$ in the case (i) and 
$8.3\times 10^6$ in the case (ii). 
The result justifies the above observation such that the model could be 
analyzed as a single field inflation during the required e-foldings number ${\cal N}_k$.

It is also found in Fig.~1 that the first several oscillations after the end of inflation 
can be well described by $\hat\chi$ since $\varphi$ takes a constant value there. 
Unfortunately, since $\varphi$ used to define the inflaton $\hat\chi$ 
is not kept constant after that as shown in the lower panels, $\hat\chi$ is not well-defined there. 
The description of preheating by using $\hat\chi$ is not justified
and each field should be treated independently in the analysis.
It should be noted that these panels also show the convergence of $\rho$ to a vacuum value
$\frac{\pi}{4}$ after a transitional period subsequent to the end of inflation.

After the end of inflation, each scalar begins oscillation starting from 
the bottom of the potential valley that is represented by eq.(\ref{valley}) 
or (\ref{valley2}) in respective cases. 
If we take account of it and the condition (\ref{ks}) for the potential (\ref{lowp}), 
each scalar is expected to oscillate satisfying 
$\sigma^2\simeq -\frac{\kappa_{\sigma S}}
{2\tilde\kappa_\sigma}(S_R^2+S_I^2)$ where $w$ and $u$ are neglected.  
It suggests that $\chi$ defined in eq.~(\ref{defx}) is found to be a good variable 
to describe the relevant background oscillation. Moreover,  
its potential might be approximated by $\frac{1}{4}\hat\kappa_S\chi^4$ 
in the case (i) and $\frac{1}{4}(\hat\kappa_S+4\alpha)\chi^4$ in the case (ii).
We call such a $\chi$ as $\chi_c$.
In Fig.~2, we plot $\chi$ which is determined by using the solutions 
of eq.~({\ref{ceofm}) at such a period. In the same panel, we also present the time 
evolution of $\chi_c$ which has the above mentioned quartic potential.
The figure shows that $\chi_c$ gives a rather good approximation for $\chi$ in 
the case (ii) where $\rho$ is kept constant throughout the oscillation period.
We also note that $\chi$ could occupy the dominant energy density of the inflaton 
$\hat\chi$ under the condition (\ref{ks}) 
since $|\chi|\gg |\sigma|$ is satisfied during inflation and then 
$\rho_\chi>\rho_\sigma$.
Thus, we could analyze the preheating by treating $\chi_c$ as a 
background oscillating field instead of the inflaton $\hat\chi$. 
On the other hand, the situation is largely different in the case (i) 
on this procedure.
Since $\rho$ starts changing the value at the oscillation period, 
$\chi=\sqrt{S_R^2+S_I^2}$ quits to reach the zero at each oscillation
after several oscillations as found from the figure.\footnote{It should be also noted
that the zero crossing of $\chi$ after $z\sim 2\times 10^7$ depends on the
value of the VEVs $w$ and $u$. The difference of $n_k$ behavior at this period 
found in Fig.~3 is considered to be caused by this reason. }
This suggests that the preheating is ineffective 
in the case (i).\footnote{It should be noted that
the fields generated through the preheating have couplings with the 
oscillating fields in the form of $\chi$ which does not have zero crossing.
We see it in the next part.} 
On the basis of this observation, in the next part we study the preheating 
and estimate reheating temperature.  

Finally, it is useful to give a comment on the domain wall problem caused by the
spontaneous $CP$ violation. Since the inflation occurs in the $CP$ violating valley 
and then the $CP$ symmetry is violated during the inflation, the expected domain 
wall is inflated away. Moreover, it is not recovered throughout the inflaton oscillation 
period since $\rho\not=0$ is kept in the present scenario as shown in the lower 
panels of Fig.~1.\footnote{ Even if different patches of the Universe take 
different $\rho$ values during the oscillation period of $\rho$ and domain walls 
happen to be produced among them, they are unstable topologically and disappear 
through the decay.}  
Thus, the problem does not appear as long as the reheating temperature is 
lower than the $CP$ breaking scale $u$.
It is noticeable that even such a low reheating temperature could make 
leptogenesis successful in the present model as discussed later.

\begin{figure}[t]
\begin{center}
\includegraphics[width=7.7cm]{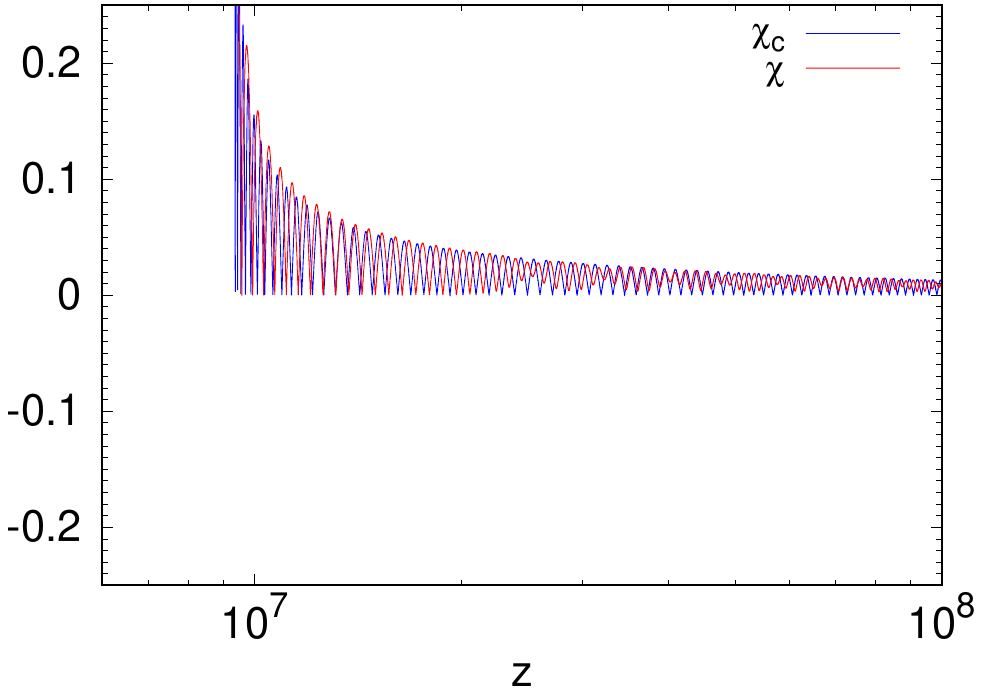}
\hspace*{5mm}
\includegraphics[width=7.5cm]{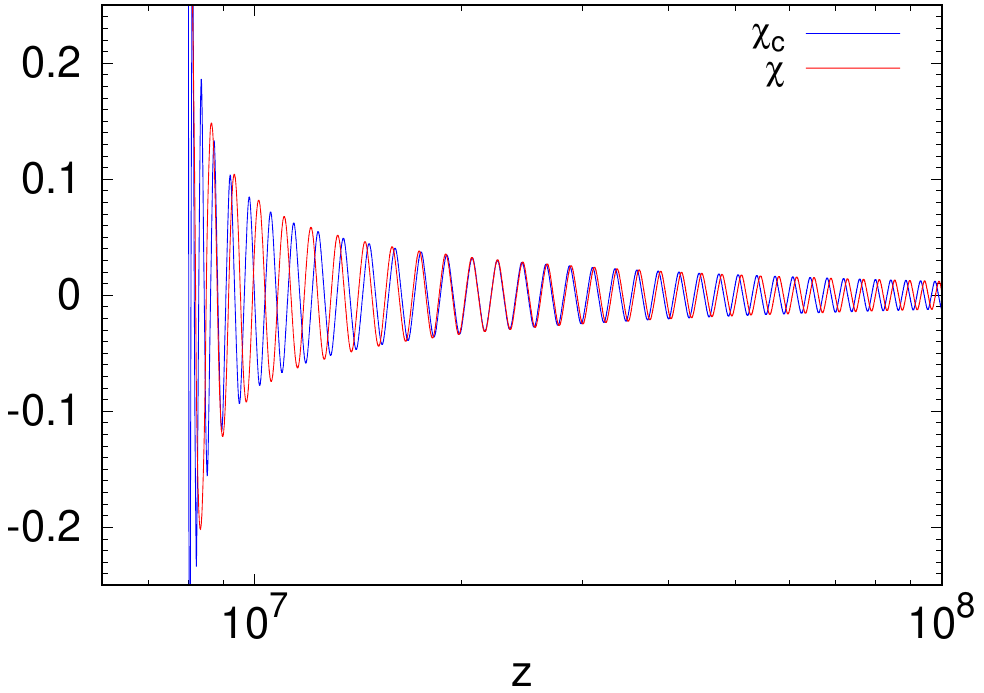}
\end{center}
\vspace{-2mm}
\footnotesize{{\bf Fig.~2}~~~~Oscillation behavior of $\chi$ determined 
by using the solutions of eq.~(\ref{ceofm}) is shown as $\chi$. 
On the other hand, one of $\chi$ which is assumed to have 
the quartic potential given in the text is shown as $\chi_c$.
Left and right panels correspond to the cases (i) and (ii), respectively.
In both cases, $w=u=0$ is assumed. }
\end{figure}

\subsection{Preheating and reheating temperature}
The background field oscillation after the end of inflation is described 
by using eq.~(\ref{oscilpot}) for $\hat\chi< M_{\rm pl}$ 
as\footnote{In the following practical 
calculation, we adopt $\chi$ as an oscillating field instead of the inflaton 
$\hat\chi$ taking account of the previous discussion.}
\begin{equation}
\frac{d^2\chi}{dt^2}+3H\frac{d\chi}{dt}+\frac{dV(\chi)}{d\chi}=0.
\label{infeq}
\end{equation}
Since the amplitude of $\chi$ evolves approximately as  
$\frac{\xi_\chi}{\sqrt{\pi\hat\kappa_S}t}$ in the quadratic 
potential, the background field $\chi$ oscillates almost 
$\frac{1}{2\pi\sqrt{3\pi}}(\xi_\chi-1)$ times before the potential 
changes from the quadratic form to the quartic one. 
This suggests that the inflaton oscillates a few times at most 
in the quadratic potential for $\xi_\chi<10$.
Thus, preheating under the quadratic potential could play 
no substantial role there and preheating in the present model 
can be studied only through the quartic potential.

Preheating under the background oscillation of $\chi$ 
generates the excitations of scalar $\chi$ itself and other scalars 
$\psi$ at its zero crossing \cite{pre}.\footnote{
We should recall that the zero crossing of the oscillating 
$\chi$ throughout this period occurs only in the case (ii).} 
If we express $\chi$ and $\psi$ with the comoving 
momentum $k$ as $\Psi_k$,
its time evolution is expressed by the equation
\begin{equation}
\frac{d^2\Psi_k}{dt^2} +3H\frac{d\Psi_k}{dt} +\omega^2_{\Psi_k}\Psi_k=0,
\label{genp}
\end{equation}
where $\omega_{\Psi_k}^2=\frac{k^2}{a^2} +g_\Psi\chi^2$ and $g_\Psi$ is a 
coupling constant between $\chi$ and $\Psi$.
Using the solution of coupled equations (\ref{infeq}) and (\ref{genp}),
the number density of produced $\Psi_k$ in the comoving frame 
can be calculated as
\begin{equation}
n_{k}=\frac{\omega_{\Psi_k}}{2}
\left(\frac{|\dot {\cal Y}_k|^2}{\omega_{\Psi_k}^2} +|{\cal Y}_k|^2\right)-\frac{1}{2},
\label{nkc}
\end{equation}
where ${\cal Y}_k=a^{3/2}\Psi_k$ and $\dot {\cal Y}_k=\frac{d{\cal Y}_k}{dt}$. 
The time evolution of $n_k$ in the present model is shown in Fig.~3 for
typical values of the VEVs of $\sigma$ and $S$. 
It is noticeable that these VEVs do not affect $n_k$ at the initial
stage of the oscillation where $n_k$ is small and then nonlinearity
in the $n_k$ evolution is negligible. 
The left panel for the case (i) shows that the particle production becomes 
ineffective after the time when the zero crossing of $\chi$ stops as expected.
On the other hand, the right panel for the case (ii) shows that the particle 
production expected in the pure quartic potential is realized at 
$z~{^>_\sim}~7\times 10^7$ which is later than the time when the potential 
changes from the quadratic one to the quartic one. 
This is considered to be brought about by the existence 
of a small quadratic term of $S_I$ \cite{confpre}, which is caused by the deviation 
of $S_R$ and $\sigma$ from their vacuum values. 
Although it could make the nature of preheating complicated in the present model, 
it does not largely affect the estimation of the reheating temperature 
as discussed later. 

\begin{figure}[t]
\begin{center}
\includegraphics[width=7.5cm]{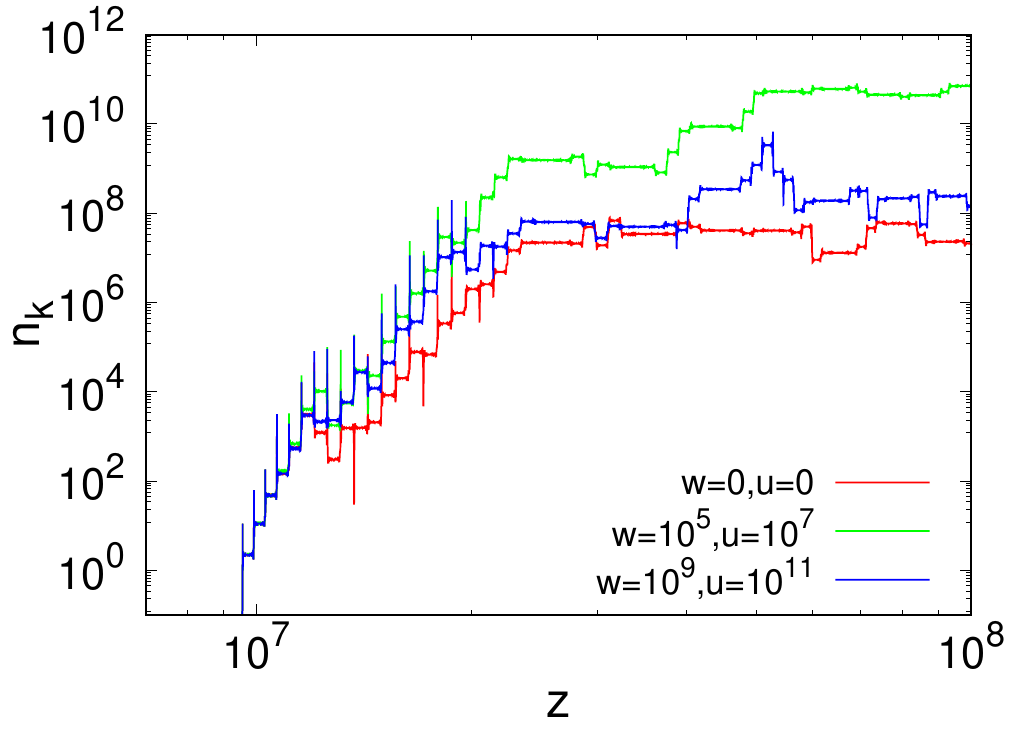}
\hspace*{5mm}
\includegraphics[width=7.5cm]{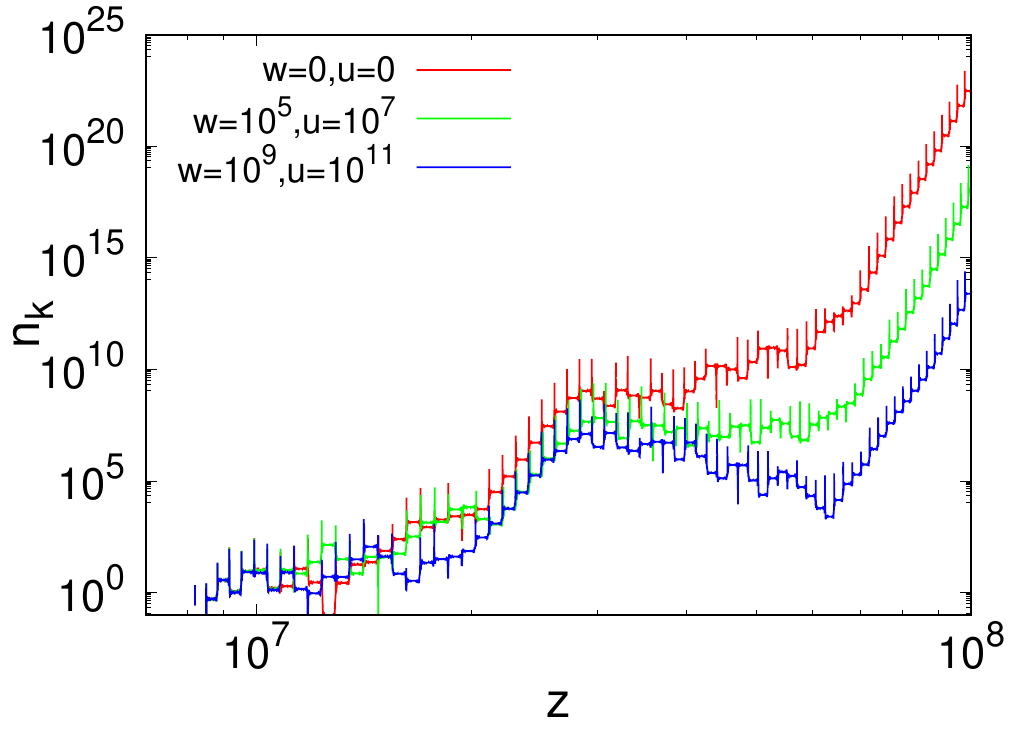}
\end{center}
\vspace{-2mm}
\footnotesize{{\bf Fig.~3}~~Time evolution of the number densisy $n_k$ of produced particles 
with a typical momentum $k$ through the preheating for  
several values of $w$ and $u$ in a GeV unit.
Left panel is for the case (i) and right panel is for the 
case (ii).  They are estimated by using eq.~(\ref{nkc}) under an assumption
that the generated particles are stable enough to be accumulated at 
every zero crossing of $\chi$.}
\end{figure}

The model with the quartic potential 
$V(\chi)=\frac{\hat\kappa_S}{4}\chi^4$ becomes conformally 
invariant \cite{confpre} and then the stochastic nature of particle 
production is known to disappear since eq.~(\ref{infeq}) reduces to the one 
in the Minkowski space. 
In order to describe it, we introduce a dimensionless conformal time 
$\tau$ and a rescaled oscillating field $f(\tau)$ which are defined by using a scale 
factor $a(\tau)$ as 
\begin{equation}
 ad\tau=\sqrt{\hat\kappa_S}\chi_0dt, \qquad
f=\frac{a\chi}{ \chi_0},
\label{resc}
\end{equation}
where $\chi_0$ is an amplitude of $\chi$ at a suitable time.
By using these variables, eq.~(\ref{infeq}) can be rewritten as
\begin{equation}
\frac{d^2f}{d\tau^2}+f^3=0.
\label{infeq1}
\end{equation}  
The solution describing this equation of motion is known to be a Jacobi 
elliptic function 
$f(\tau)={\rm cn} \left(\tau,\frac{1}{\sqrt 2}\right)$ \cite{confpre}.
If we use the Friedman equation for this oscillation of the field
which dominates the inflaton energy, we find 
\begin{equation}
a(\tau)=\frac{\chi_0}{2\sqrt3 M_{\rm pl}}\tau, \qquad  
\tau=2(3\hat\kappa_S M_{\rm pl}^2)^{1/4}\sqrt t.
\label{rd}
\end{equation} 
Since the Hubble parameter behaves as $H=\frac{1}{2t}$ under eq.~(\ref{rd}), 
this oscillation era is found to be described as the radiation dominated one.
It means that the radiation domination starts 
soon after the end of inflation. 
It is a distinguished feature of the preheating in the case (ii). 

In order to estimate the energy transfer of the inflaton to excitations of 
the scalars $S_R$, $S_I$ and $\sigma$, we study it by noting their 
interaction with the oscillating field $\chi$.
The interaction among these scalars are expressed by the potential (\ref{lowp}).
It shows that the frequency of the oscillating field $\chi$
is approximated as $\sqrt{\hat\kappa_S}|\chi|$, 
which could be much larger than the one of $\sigma$.   
On the other hand, the mass eigenvalues of the scalars caused 
by the interaction with the oscillating field $\chi$ are found 
to be approximated as 
\begin{eqnarray}
 3\hat\kappa_S\chi^2, \quad
 \hat\kappa_S\chi^2, \quad |\kappa_{\sigma S}|\chi^2.
\label{imass2}
\end{eqnarray}
These values suggest that the transfer of the oscillation energy of $\chi$ 
to these mass eigenstates is kinematically forbidden for the parameters given 
in (\ref{para}).

Here, we take account of the interactions of $S$ and $\sigma$ 
with the Higgs scalar $\phi$ which are neglected by now as 
\begin{equation}
\kappa_{\phi S}S^\dagger S\phi^\dagger\phi+ 
\frac{1}{2}\kappa_{\phi\sigma}\sigma^2\phi^\dagger\phi,
\label{oscm}
\end{equation} 
where $S$ number violating terms are assumed to be zero.
They cause the mass for $\phi$ as
\begin{equation}
m_\phi^2\simeq\frac{1}{2}\left(\kappa_{\phi S}\chi^2+\kappa_{\phi\sigma}\sigma^2\right).
\label{imass1}
\end{equation}
It shows again that the decay of the oscillation of $\chi$ to $\phi$ is 
kinematically forbidden unless $\kappa_{\phi S}$ and 
$\frac{|\kappa_{\sigma S}|}{2\tilde\kappa_\sigma}\kappa_{\phi\sigma}$ is smaller 
than $\hat\kappa_S$.\footnote{Even if $S$ number violating terms are introduced,
this conclusion does not change as long as the similar conditions 
for the coupling constants are satisfied.}
On the other hand, such small $\kappa_{\phi S}$ and $\kappa_{\phi\sigma}$ cannot 
cause the effective decay of the $\chi$ oscillation. 
As a result, the energy transfer from the oscillation of $\chi$ 
to the scalar excitations is expected to proceed only at the time when it crosses 
the zero where the nonperturbative particle production could occur.

Preheating under the background oscillation of $\chi$ can 
generate  excitations of the scalar $\chi$ itself and other scalars 
$\psi$ at its zero crossing \cite{pre}.\footnote{
We should recall that the zero crossing of the oscillating 
$\chi$ throughout this period occurs only in the case (ii), where 
$\psi$ stands for $S_R, \sigma$, and $\phi$.} 
In a quartic potential case \cite{confpre}, the model becomes 
conformally invariant and  has no stochastic nature as mentioned above. 
Time evolution equations for $\chi_k$ 
and $\psi_k$, which are the modes with a comoving momentum $k$,  
can be transformed to simple ones by rescaling them 
to dimensionless quantities. 
If we define rescaled variables as
\begin{eqnarray}
X_k=\frac{a\chi_k}{\chi_0 }, \qquad 
F_k=\frac{a\psi_k}{\chi_0 }, \qquad 
\bar k=\frac{k}{\chi_0\sqrt{\hat\kappa_S}},
\end{eqnarray}
the time evolution equations for them are given as 
\begin{eqnarray}
&& X_k^{\prime\prime}+\bar\omega_{X_k}^2X_k=0, \qquad 
\bar\omega_{X_k}^2=\bar k^2+3f(\tau)^2, \nonumber \\
&& F_k^{\prime\prime}+\bar\omega_{F_k}^2F_k=0, \qquad 
\bar\omega_{F_k}^2=\bar k^2+\frac{g_\psi}{\hat\kappa_S}f(\tau)^2, 
\label{teq}
\end{eqnarray}
where the prime stands for a $\tau$ derivative and the function $f(\tau)$ 
is the solution of eq.~(\ref{infeq1}).
$g_\psi$ is a coupling constant of the scalar $\psi$ with the oscillating 
scalar $\chi$ and they can be read off from eqs.~(\ref{imass2}) 
and (\ref{imass1}).\footnote{We note that $g_\chi$ is expressed 
as $g_\chi=3\hat\kappa_S$.}  
The number density of produced particles with the momentum $k$ 
can be calculated by using the solution of eq.~(\ref{teq}) as
\begin{equation}
\bar n_{\bar\Psi_k}=\frac{\bar\omega_{\bar\Psi_k}}{2}
\left(\frac{|\bar\Psi_k^\prime|^2}{\bar\omega_{\bar\Psi_k}^2} 
+|\bar\Psi_k|^2\right)-\frac{1}{2},
\label{nk0}
\end{equation}
where $\bar\Psi_k$ represents both $F_k$ and $X_k$.

Analytic estimation of $\bar n_{\bar\Psi_k}$ has been studied in \cite{confpre}. 
Since an adiabaticity condition 
$\bar\omega_{\bar\Psi_k}^\prime<\bar\omega_{\bar\Psi_k}^2$ 
could be violated for momentum modes with 
$\bar k^2< \frac{1}{3}\sqrt{\frac{g_\Psi}{\hat\kappa_S}}$ 
at the zero crossing of $\chi$, excitations of $\bar\Psi_k$ with a momentum  $k$ in
such a region are expected to be produced.
If the produced particles are accumulated at each zero crossing of $\chi$,
$\bar n_{\bar\Psi_k}$ is shown to have an exponential behavior $e^{2\mu_k\tau}$ 
where the exponent $\mu_k$ is characterized by $\frac{g_\Psi}{\hat\kappa_S}$.
However, if the produced particles decay immediately much before the next 
zero crossing of $\chi$, $\bar n_{\bar\Psi_k}$
does not show such a behavior since the produced excitations 
do not affect the particle production at the next zero crossing.
Moreover, the $\psi$ decay cannot transfer large energy effectively 
from the oscillation to radiation
if $\psi$ decays soon after its creation time much before the $\chi$ 
amplitude becomes large to result in large $\psi$ mass. 

By using an approximated analytic solution of eq.~(\ref{teq}) and
neglecting the accumulated particle effect, 
momentum distribution of the produced $\bar\Psi_k$ 
through one zero crossing of $\chi$ can be estimated as \cite{confpre}
\begin{equation}
\bar n_{\bar\Psi_k}=e^{-\left(\bar k/\bar k_c\right)^2},
\qquad \bar k_c^2=\sqrt\frac{g_\Psi}{2\pi^2\hat\kappa_S},
\label{nk}
\end{equation}
where $\bar n_{\bar\Psi_k}=0$ is assumed before each zero crossing of $\chi$.
By using this result, the number density of the produced $\bar\Psi$ during 
one zero crossing is found to be 
\begin{equation}
\bar n_{\bar\Psi}=\int\frac{d^3\bar k }{(2\pi)^3}\bar n_{\bar\Psi_k}
=\int\frac{d^3\bar k}{(2\pi)^3}e^{-(\bar k/\bar k_c)^2}
=\frac{\bar k_c^3}{8\pi^{3/2}}.
\label{nk1}
\end{equation}
This analytic result is derived under an assumption that 
the potential of $\chi$ takes a quartic form, which is based on the 
observation on Fig.~2.
In order to confirm that the use of this formula could be justified 
in the present case, we solve 
the coupled equations for $\chi$ and $\Psi$ which correspond to 
eqs.~(\ref{infeq})  and (\ref{genp}) to estimate eq.~(\ref{nk0}) 
by using the solutions. The result is given in Fig.~4. 
We use $\chi_0=0.1M_{\rm pl}$
for the analytic estimation of $\bar n_{\bar\Psi_k}$.
It is consistent with the fact that quartic 
potential starts at the time when the amplitude of $\chi$ 
becomes $\frac{M_{\rm pl}}{\xi_\chi}$. 
From this figure, the present analytic estimation is found 
to give a rather good approximation for $\bar n_{\bar\Psi_k}$. 

\begin{figure}[t]
\begin{center}
\includegraphics[width=7.5cm]{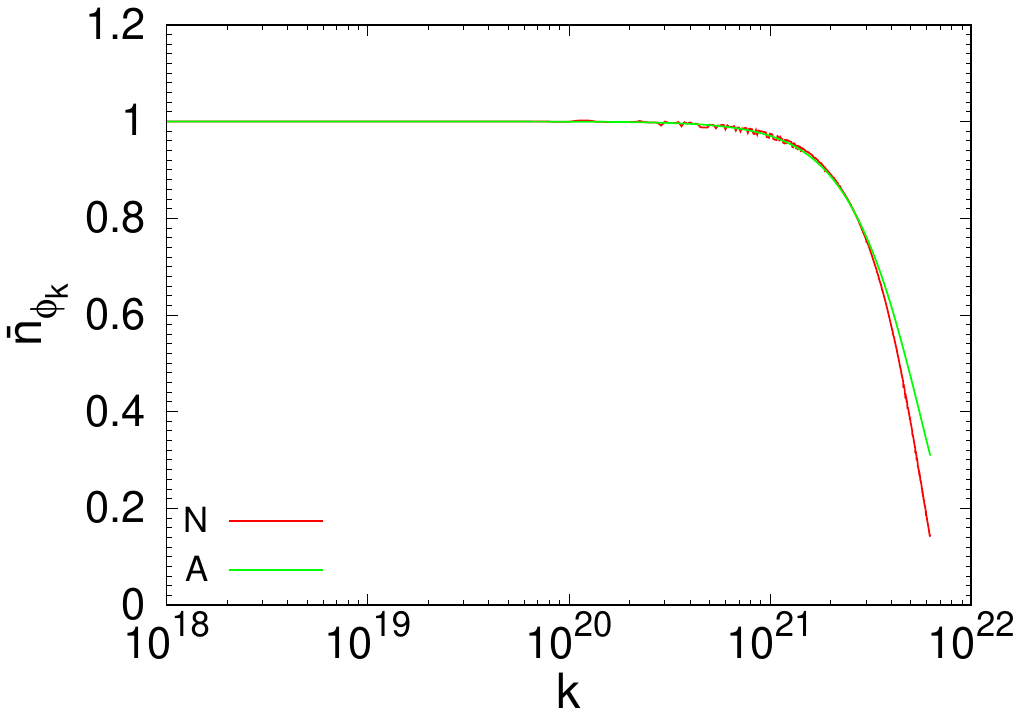} 
\end{center}
\vspace{-2mm}
\footnotesize{{\bf Fig.~4}~~Momentum distribution of the produced 
Higgs excitations for the case (ii) with $g_\Psi=10^{-5}$ and $\chi_0=0.1M_{\rm pl}$. 
Analytic results and numerical ones are plotted for a nonrescaled 
momentum $k$ in the comoving frame as A and N, respectively.}
\end{figure}

Hereafter we focus on a case $\Psi=\phi$ with $g_\phi/\hat\kappa_S> 1$,
which is expected to give a dominant contribution.\footnote{Other cases are 
briefly discussed in Appendix A.}
Energy transfer from the $\chi$ oscillation to relativistic particles
is caused through the decay of the produced Higgs scalar $\phi$ to the
light SM fermions. We note that the mass of the SM contents 
except for $\phi$ is irrelevant to $\chi$ and then they are kept light 
throughout the relevant process. 
The decay of the produced $\phi$ mainly proceeds through 
$\phi\rightarrow \bar qt$ which is caused by a large top Yukawa coupling $h_t$. 
The decay width in the comoving frame is given by using the conformally 
rescaled unit as
\begin{eqnarray}
&&\bar\Gamma_\phi=\frac{3 h_t^2}{8\pi}\bar m_\phi,  \qquad 
\bar m_\phi=\frac{a m_\phi}{ \chi_0 \sqrt{\hat\kappa_S}}
=\sqrt\frac{g_\phi}{\hat\kappa_S}f(\tau).
\end{eqnarray} 
Since $\bar\Gamma_{\phi}^{-1}<\tau_0/2$ is satisfied 
in the case $g_\phi> 4\times 
10^{-7}\left(\frac{\hat\kappa_S}{10^{-8}}\right)$
for the $\chi$ oscillation period $\tau_0=7.416$, the produced $\phi$
could decay through this process completely before the next 
zero crossing of $\chi$. In such a case, since the produced $\phi$ 
is not accumulated, the $\phi$ production at the next zero crossing 
has no nonlinear effect caused by it.
Thus, the previous result (\ref{nk}) is applicable.

If we fix $\tau=0$ at the time of the first zero-crossing, $f(\tau)$ can be
expressed approximately as $f(\tau)=\chi_0\sin(2\pi \chi_0\frac{\tau}{\tau_0})$.
Therefore, the energy density transfered to radiation in the comoving frame 
through the $\phi$ decay during a half period of the $\chi$ oscillation 
can be estimated as
\begin{equation}
\delta\bar\rho_r=\int^{\tau_0/2}_0d\tau\bar\Gamma_\phi\bar m_\phi\bar n_\phi
e^{-\int_0^\tau\bar\Gamma_\phi\tau^\prime}
=\frac{1}{8\pi^{3/2}(2\pi^2)^{3/4}}\left(\frac{g_\phi}{\hat\kappa_S}\right)^{5/4}
Y(\chi_0,\gamma_\phi),
\end{equation}
where $\gamma_\phi$ and $Y(\chi_0,\gamma_\phi)$ are defined as
\begin{equation}
\gamma_\phi=\frac{3 h_t^2\tau_0}{16\pi^2}\sqrt\frac{g_\phi}{\hat\kappa_S}, \qquad
Y(\chi_0,\gamma_\phi)=\frac{2\pi\gamma_\phi}{\tau_0}
\int^{\tau_0/2}_0d\tau \chi_0^2\sin^2(2\pi \chi_0\frac{\tau}{\tau_0})
e^{-2\gamma_\phi\sin^2(\frac{\pi \chi_0\tau}{\tau_0})}.
\end{equation}
The energy converted to radiation is accumulated linearly 
at each $\chi$ zero crossing and then its averaged density during a period 
$\tau$ is estimated as
\begin{equation}
\bar\rho_r(\tau)=\frac{2\tau}{\tau_0}\delta\bar\rho_r
=6.5\times 10^{-4}\left(\frac{g_\phi}{\hat\kappa_S}\right)^{5/4}
Y(\chi_0,\gamma_\phi)\tau,
\label{rde}
\end{equation}
where substantial change of the amplitude $\chi_0$ is assumed to be 
negligible.

Since the total energy density of the $\chi$ oscillation energy $\bar\rho_\chi$
and the transfered energy to the radiation $\bar\rho_r$   
is conserved, reheating temperature realized through
this process can be estimated from 
$\bar\rho_r=\bar\rho_{\chi_0}$. 
If we rewrite this relation by using the physical unit, we have
\begin{equation}
\frac{\hat\kappa_S}{4}\left(\frac{\chi_0}{a}\right)^4=\frac{\pi^2}{30}g_\ast T_R^4.
\end{equation}
By applying eqs.~(\ref{rd}) and (\ref{rde}) to this formula, 
the reheating temperature is found to be
\begin{equation}
T_R=5.9\times 10^{15}g_\phi^{5/4}Y(\chi_0,\gamma_\phi)~{\rm GeV},
\label{nonptr}
\end{equation}
where we use $g_\ast=130$ in this model.
The previously mentioned complex feature 
of the preheating is found to be irrelevant to this estimation of the reheating 
temperature.

If the energy transfer to $\phi$ by preheating is not effective, it does not play a 
substantial role for reheating. The dominant energy continues to be kept 
in the oscillation of $\chi$ and reheating is expected to proceed through 
perturbative processes which occur when the oscillation amplitude of $\chi$ becomes $O(u)$.
We note that the case (i) corresponds to this class.  
Even if the preheating is effective, it cannot complete the energy transfer from 
inflaton since it occurs only at a period where inflaton amplitude is large enough. 
Thus, the reheating is finished through the perturbative process.
In the case (ii) both of them could play a role to determine reheating temperature.

If assumed coupling constants allow $2m_\phi<m_\chi$, 
the $\chi$ decay occurs mainly through 
$\chi \rightarrow \phi^\dagger\phi$ at tree level.  
Its decay width can be estimated as
\begin{eqnarray}
\Gamma_\chi&\simeq& 
\frac{g_\phi^2}{16\pi \hat\kappa_S}m_\chi\sqrt{1-\frac{4m_\phi^2}{m_\chi^2}}. 
\label{width}
\end{eqnarray}
After $\chi$ decays to a $\phi$ pair,
the SM contents are expected to be thermalized through the SM interactions 
immediately. 
If $\Gamma_\chi\ge H$ is satisfied at $|\chi|\simeq u$ which requires
$g_\phi>10^{-7.1}\left(\frac{\hat\kappa_S}{10^{-8}}\right)^{1/2}
\left(\frac{u}{10^{11}~{\rm GeV}}\right)^{1/2}$,
reheating temperature can be estimated by using 
$\frac{1}{4}\hat\kappa_Su^4=\frac{\pi^2}{30}g_\ast T_R^4$ as
\begin{equation}
T_R\simeq
2.8\times10^8\left(\frac{\hat\kappa_S}{10^{-8}}\right)^{1/4}
\left(\frac{u}{10^{11}~{\rm GeV}}\right)~{\rm GeV}.
\end{equation}
We note that this reheating temperature depends on $u$ but 
not on $g_\phi$ in contrast with the one of preheating. 
On the other hand, if the coupling $g_\phi$ does not satisfy the above condition,
the instantaneous decay does not occur and the reheating temperature 
should be estimated from the condition $\Gamma_\chi\simeq H$. 

\begin{figure}[t]
\begin{center}
\includegraphics[width=7.7cm]{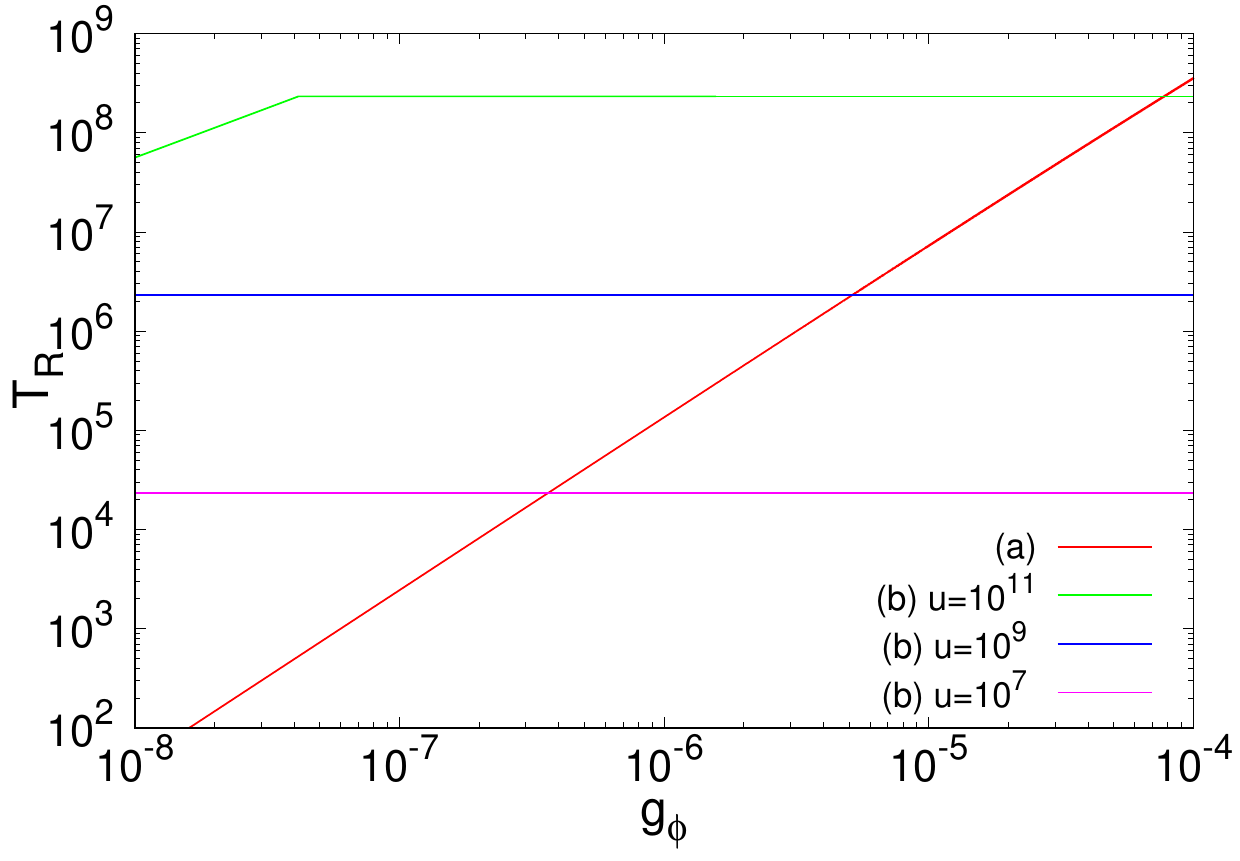}
\end{center}
\vspace{-2mm}
\footnotesize{{\bf Fig.~5}~~(a) Reheating temperature as a function of the
coupling $g_\phi$ predicted by the preheating. In this estimation 
$\chi_0=0.1M_{\rm pl}$ is assumed.
(b) Reheating temperature predicted by the perturbative process for some
values of the VEV $u$ in a GeV unit.}
\end{figure}

In Fig.~5, using the parameters given in eq.~(\ref{para}),
 the expected reheating temperature through 
both processes are plotted for several values of $u$ as a function of 
$g_\phi(=\kappa_{\phi S})$. 
In the case (i) reheating temperature is fixed independently of the coupling $g_\phi$
since reheating occurs through the perturbative process.
On the other hand, in the case (ii) the figure shows the reheating temperature 
is determined by the preheating at larger values of $g_\phi$ but it is fixed by 
the perturbative process at smaller ones. It crucially depends on a value of $u$. 
If we take account of constraints from perturbativity 
of the model, $g_\phi<10^{-4.4}$ is required \cite{cpinf}.
Thus, in both cases the reheating temperature cannot be 
higher than $10^{8}$~GeV for $u~{^<_\sim}~10^{11}$ GeV 
as found from the figure.
In the present scenario, we find that the energy transfer from the $\chi$ 
oscillation to the radiation due to the preheating cannot be efficient 
to realize a higher reheating temperature compared with the 
perturbative one since the decay of the 
produced $\phi$ is so effective. 

\begin{figure}[t]
\begin{center}
\begin{tabular}{c|cccc}\hline
case & $T_R$ & ${\cal N}_{k_\ast}$ & $n_s$ & $r$ \\ \hline
    & $10^4$ & 46.87 & 0.9552 & 0.00546  \\ 
(i) & $10^6$ & 48.40&  0.9568&  0.00512  \\
    & $10^8$ & 49.92 & 0.9581 & 0.00481  \\ \hline
(ii) &            & 54.16& 0.9615 & 0.00409 \\ \hline  
\end{tabular}
\end{center}
\footnotesize{{\bf Table 1}~~ Predicted values of the e-foldings number ${\cal N}_{k_\ast}$,
the scalar spectral index $n_s$ and the tensor-to-scalar ratio $r$ at the pivot scale
$k_\ast=0.05$ Mpc$^{-1}$ for the cases (i) and (ii). In the case (ii), ${\cal N}_{k_\ast}$ has no
dependence on the reheating temperature since radiation domination starts soon after 
the end of inflation. Reheating temperature is shown in a GeV unit.}
\end{figure}

Finally, we consider a e-foldings number ${\cal N}_{k_\ast}$ at a pivot scale $k_\ast$
by taking account of the reheating temperature analyzed above. 
Since the reheating temperature 
is predicted, we can assign a specific value of ${\cal N}_{k_\ast}$ by using it.
If we note a relation $k=a_kH_k$ with $H_k^2=\frac{V_k}{3M_{\rm pl}^2}$, 
it can be calculated as 
\begin{equation}
{\cal N}_{k_\ast}=55.43-\ln\left(\frac{k_\ast}{a_0H_0}\right)
+\frac{1}{3}\ln\left(\frac{T_R}{10^8~{\rm GeV}}\right)
+\frac{2}{3}\ln\left(\frac{M_{k_\ast}}{10^{16}~{\rm GeV}}\right),
\label{casei}
\end{equation}
where $V_{k_\ast}=M_{k_\ast}^4$ and $g_\ast=130$ is used. 
On the other hand, if radiation domination starts soon after the
end of inflation as in the case (ii), reheating temperature dependence of ${\cal N}_{k_\ast}$
is lost and it is found to expressed as  
\begin{equation}
{\cal N}_{k_\ast}=59.76-\ln\left(\frac{k_\ast}{a_0H_0}\right)
+\ln\left(\frac{M_{k_\ast}}{10^{16}~{\rm GeV}}\right).
\label{caseii}
\end{equation}
If we use these relations and eq.~(\ref{power}), we could discuss a favorable reheating 
temperature on basis of future CMB data. In fact, 
since eq.~(\ref{power}) and an $A_s$ value given in \cite{planck18} fix $M_{k_\ast}$ to
$M_{k_\ast}=6.01\times 10^{16}\frac{1}{\sqrt{{\cal N}_{k_\ast}}}$,  
${\cal N}_{k_\ast}$ can be determined by using it in eq.~(\ref{casei}) or (\ref{caseii}).
Moreover, if we note that the scalar spectral index $n_s$ and the tensor-to-scalar ratio $r$
are expressed by using the slow roll parameters $\epsilon$ and $\eta$ as
$n_s=1-6\epsilon +2\eta$ and $r=16\epsilon$, 
they can be predicted for a specific value of ${\cal N}_{k_\ast}$ because of the relations 
given in eq.~(\ref{slowroll}). Their predicted values are shown for 
$k_\ast=0.05$ Mpc$^{-1}$ in Table 1.

\section{Relevant phenomenology}
This inflation scenario can be closely related to some low energy phenomena
in an interesting way if the model consists of additional fields. 
As such candidates, here we consider  down-type vectorlike quarks 
$D_{L,R}$\footnote{ They have the same transformation property as the ordinary 
right-handed down-type quarks $d_{R_j}$ for the SM gauge groups.}, 
three right-handed neutrinos $N_j$ and an inert doublet scalar $\eta$. 
Their $Z_4\times Z_4^\prime$ charges are 
assigned as $D_L (2,0), D_R (0, 2), N_j (1,1)$ and $\eta (-1,-1)$.
After $S$ and $\sigma$ get the VEVs, the neutrino sector is reduced to the 
scotogenic neutrino mass model with $Z_2$ invariance, which can explain the 
small neutrino mass and the existence of dark matter \cite{scot1,scot2}. 
Since $\eta$ has an odd parity of the remnant $Z_2$ symmetry, 
its lightest neutral component is stable 
and can be a good dark matter candidate at TeV scales \cite{ks1,ks2}. 
Neutrino mass generation in this model is addressed in relation to leptogenesis later.   
The down-type quarks $D_{L,R}$ have invariant Yukawa couplings
\begin{equation}
-{\cal L}_Y= \sum_{i,j=1}^3h_{ij}\tilde\phi\bar d_{L_i}d_{R_j}+
\sum_{j=1}^3\Big(y_{d_j}S\bar D_L d_{R_j} + \tilde y_{d_j}S^\dagger\bar D_L d_{R_j} + 
y_D\sigma\bar D_LD_R +{\rm h.c.}\Big),
\label{dyuk}
\end{equation}
where all couplings are supposed to be real and positive as a result of $CP$ invariance 
of the model. It also imposes $\theta=0$ on the $\theta$-term in the QCD sector.
During the $\chi$ oscillating period, these Yukawa couplings 
induce the mass for $D_{L,R}$ such as 
\begin{equation}
M_{D}\simeq \frac{|\chi|}{\sqrt 2}\left[\sum_{j=1}^3(y_{d_j}^2+\tilde y_{d_j}^2) +
y_D^2\frac{\kappa_{\sigma S}^2}
{4\tilde\kappa_\sigma^2}\right]^{\frac{1}{2}}.
\end{equation}
If $\hat\kappa_S<y_{d_j},~\tilde y_{d_j}$ is satisfied, the existence of these vectorlike 
fermions does not affect the previous preheating analysis because of 
 $M_D\gg \sqrt{\hat\kappa_S}|\chi|$. On the other hand, they could cause noticeable effects 
on low energy phenomena.
As such examples, we discuss a $CP$ phase in the CKM 
matrix and low scale leptogenesis.

\subsection{A $CP$ phase in the CKM matrix } 
After the electroweak symmetry breaking, 
the Yukawa couplings in eq.~(\ref{dyuk}) induce a $4\times 4$ mass matrix 
${\cal M}$ 
for down-type quarks as
\begin{equation}
(\bar d_{Li}, \bar D_L)\left(\begin{array}{cc}
{\cal M}_{d_{ij}} & 0 \\ 
{\cal F}_{d_j}  & \mu_D\\
\end{array}\right)
\left(\begin{array}{c} d_{R_j} \\ D_R \\ \end{array} \right)+{\rm h.c.},
\label{dmass}
\end{equation}
where ${{\cal M}_d}_{ij}=\frac{1}{\sqrt 2}h_{ij}v$, 
${{\cal F}_d}_j=\frac{1}{\sqrt 2}(y_{d_j}e^{i\rho}+\tilde y_{d_j}e^{-i\rho})u$ and $\mu_D=y_Dw$. 
The VEV of the ordinary Higgs doublet scalar $\phi$ is represented by $v$. 
Study of this mass matrix clarifies the nature of the $CP$ violation in the strong 
and weak interaction sectors as follows.  

If we make quark mass matrices real by using the chiral transformation 
for quarks including $D_{L,R}$,
the QCD parameter $\theta$ is shifted to $\bar\theta\equiv\theta+
{\rm arg (det}({\cal M}_u{\cal M}))={\rm arg (det}{\cal M})$ where 
$\theta=0$ and ${\rm arg(det}{\cal M}_u)=0$ are taken into account.
Since ${\rm arg(det}{\cal M})=0$ is satisfied as found from eq.~(\ref{dmass}),  
$\bar\theta=0$  is realized and then  
the strong $CP$ problem does not appear at tree level.
It is based on the Nelson-Barr mechanism \cite{nb1,nb2,nb3,bbp}.
On the other hand, if we integrate out $D_{L,R}$ in these mass terms,
we find that effective Yukawa couplings in the ordinary light quark sector could 
be complex. This suggests that $\bar\theta=0$ is realized but a $CP$ phase in 
the CKM matrix could be induced.

The CKM matrix is given by $V^{\rm CKM}=O_u^TA$ where $O_u$ is an orthogonal 
matrix used for the diagonalization of an up-type quark mass matrix ${\cal M}_u$.
$A$ is a $3\times 3$ part of the unitary matrix to diagonalize ${\cal M}$.
It can be fixed so as to diagonalize a matrix ${\cal M}{\cal M}^\dagger$ 
through 
\begin{equation}
\left(\begin{array}{cc} A & B \\ C& D \\\end{array}\right)
\left(\begin{array}{cc} {\cal M}_d{\cal M}_d^\dagger & {\cal M}_d{\cal F}_d^\dagger \\ 
  {\cal F}_d{\cal M}_d^\dagger & \mu_D\mu_D^\dagger +{\cal F}_d{\cal F}_d^\dagger \\
\end{array}\right)
\left(\begin{array}{cc} A^\dagger & C^\dagger \\ B^\dagger 
& D^\dagger \\\end{array}\right)=
\left(\begin{array}{cc} \tilde{\cal M}_d^2 & 0 \\ 0 & M_D^2 \\\end{array}\right),
\label{ddmass}
\end{equation}
where $\tilde{\cal M}_d$ is a diagonal $3\times 3$ light quark mass matrix.
Eq.~(\ref{ddmass}) requires
\begin{eqnarray}
   && {\cal M}_d{\cal M}_d^\dagger=A^\dagger \tilde{\cal M}_d^2A+ 
C^\dagger M_D^2C, \qquad
  {\cal F}_d{\cal M}_d^\dagger=B^\dagger \tilde{\cal M}_d^2A+ 
D^\dagger M_D^2C, \nonumber \\
   && \mu_D\mu_D^\dagger+{\cal F}_d{\cal F}_d^\dagger=
B^\dagger \tilde{\cal M}_d^2 B+ D^\dagger M_D^2 D.
\end{eqnarray}
Solving these conditions, $A$ is found to be fixed by \cite{ps1,ps2}
\begin{equation}
A^{-1}\tilde{\cal M}_d^2A\simeq {\cal M}_d{\cal M}_d^\dagger -\frac{1}{\mu_D^2
+{\cal F}_d{\cal F}_d^\dagger}
({\cal M}_d{\cal F}_d^\dagger)({\cal F}_d{\cal M}_d^\dagger).
\label{ckm}
\end{equation}
If both $y_{d_j}\not=\tilde y_{d_j}$ and $\mu_D^2<{\cal F}_d{\cal F}_d^\dagger$ are satisfied, 
a complex phase of $A$ could have a substantial magnitude since the second term in 
the right-hand side of eq.~(\ref{ckm}) is comparable to the first term.  
Such a situation could be realized
if the coupling constants and the VEVs satisfy the conditions \cite{cpinf} 
\begin{equation}  
\hat\kappa_S< y_{d_i} \sim \tilde y_{d_i} <y_D, \qquad  v< w < u.
\label{rehpara}
\end{equation}
In that case, scalars $S$ and $\sigma$ can give origin of $CP$ violation 
in the SM escaping the strong $CP$ problem at tree level at least. 
Concrete examples of the CKM matrix in this scenario are given in Appendix B.
These examples suggest that the present scenario 
for the spontaneous $CP$ violation could generate a realistic CKM 
matrix by assuming an appropriate texture for ${\cal M}_d$.

In this model, unitarity of the CKM matrix is violated and also 
flavor changing neutral processes (FCNCs) could appear at tree level. 
They could give a constraint on the lower bounds for values of $w$ and $u$. 
Since $\mu_D\mu_D^\dagger+{\cal F}_d{\cal F}_d^\dagger$ is assumed to be 
much larger than each component of ${\cal F}_d{\cal M}_d^\dagger$ which suggests 
$u, w\gg v$, $B, C$ and $D$ are found to be approximately expressed as
\begin{equation}
  B\simeq -\frac{A{\cal M}_d{\cal F}_d^\dagger}{\mu_D\mu_D^\dagger
+{\cal F}_d{\cal F}_d^\dagger},
 \qquad C\simeq\frac{{\cal F}_d {\cal M}_d^\dagger}{\mu_D\mu_D^\dagger
+{\cal F}_d{\cal F}_d^\dagger},
   \qquad D\simeq 1.
\end{equation}
For example,  unitarity violation in the $j$-th low of $V^{\rm CKM}$ is found to 
be described by the component $B_j$.
If we suppose that $A$ is almost diagonal, $B_j$ is roughly estimated as 
$B_j\sim \frac{h_{3j}v}{y_{d_j}u}$ as long as $ y_{d_j}$ and $\tilde y_{d_j}$ have the same 
order values.\footnote{We note that the examples given in Appendix B 
correspond to such a case.} 
On the other hand, since experimental results on the flavor changing 
neutral current processes require \cite{ubckm1,ubckm2} 
\begin{equation}
|B_1B_2^\ast|~{^<_\sim}(0.3-0.7)\times 10^{-5}\left(\frac{1~{\rm TeV}}{M_D}\right), \quad 
|B_3^\ast B_1|<(0.4-2.2)\times 10^{-4}\left(\frac{1~{\rm TeV}}{M_D}\right),
\end{equation}
$u>10^4$ GeV should be satisfied.
The $Z$ decay to hadrons also gives  a constraint $|B_1|^2<1.7\times 10^{-3}$ because of
the existence of off-diagonal couplings of $Z$ with the down-type quarks.
It is interesting that the unitarity violation in the CKM matrix or a certain 
FCNC might be found in future precise experiments. 
If such a signature is confirmed there, it might be a hint for the present 
inflation scenario in which $u$ takes a value near the above mentioned 
lower bound.
 In the next subsection, we show that even such a small value of $u$ could allow 
the successful leptogenesis in the present scenario.

Finally, we should also note that a Dirac $CP$ phase in the PMNS matrix \cite{pmns1,pmns2} 
could be also explained in the similar way as the CKM matrix.
In fact, if vectorlike charged leptons are introduced in the model and 
neutrinos are assumed to have large flavor mixing, 
the same type mass matrix of charged leptons as eq.~(\ref{dmass}) could make it possible. 
We will discuss it elsewhere.

\subsection{Low scale leptogenesis}
Small neutrino mass is usually considered to be generated through the Weinberg operator
\begin{equation}
\sum_{\alpha,\beta=1}^3\frac{f_{\alpha\beta}}{\Lambda}\bar\ell_{L_\alpha}{\phi}
\bar\ell_{L_\beta}\phi,
\end{equation}
where $\Lambda$ is a cutoff, and $\ell_{L_{\alpha}}$ and $\phi$ are 
the left-handed doublet lepton and the doublet Higgs scalar, respectively. 
If we take account of neutrino oscillation data \cite{pdg}, we find that
$\frac{f_{\alpha\beta}}{\Lambda}=O(10^{-11})~{\rm GeV}^{-1}$ is 
required to explain them.
In the type I seesaw scenario \cite{seesaw1,seesaw2,seesaw3}, 
the neutrino Yukawa coupling 
$h_{\alpha j}$ and the right-handed neutrino mass $M_{N_j}$ determine the coupling 
$\frac{f_{\alpha\beta}}{\Lambda}$ as $\sum_j\frac{h_{\alpha j}h_{\beta j}}{M_{N_j}}$. 
In this case,  baryon number asymmetry is known to be generated 
through thermal leptogenesis \cite{leptg} via the decay of the lightest right-handed neutrino $N_1$.
$CP$ asymmetry in this decay is given by \cite{cpasym1,cpasym2,cpasym3}
\begin{equation}
\varepsilon=\frac{1}{8\pi}\sum_{j=2,3}
\frac{{\rm Im}[ \sum_{\alpha}(h_{\alpha 1}
h_{\alpha j}^\ast)]^2}
{\sum_{\alpha} h_{\alpha 1}h_{\alpha 1}^\ast}
F\left(\frac{M_{N_j}^2}{M_{N_1}^2}\right),
\end{equation}
where $F(x)=\sqrt x\left[1-(1+x)\ln\frac{1+x}{x}\right]$.
If we require that successful leptogenesis occurs consistently with the neutrino 
oscillation data
in this scenario, $M_{N_1}$ should be larger than $10^9$ GeV \cite{di} 
as long as there is no other process to generate the right-handed neutrinos 
than the one caused by the neutrino Yukawa couplings $h_{\alpha j}$. 

In radiative neutrino mass scenarios, $\frac{f_{\alpha\beta}}{\Lambda}$ is
fixed in a different way from the above type I seesaw scenario.
For example, in a scotogenic model \cite{scot1,scot2} which has neutrino Yukawa 
couplings $h_{\alpha j}\bar\ell_\alpha\eta N_j$ with an inert doublet scalar $\eta$, 
a small scalar coupling $\frac{\lambda_5}{2}[(\phi^\dagger\eta)^2+ {\rm h.c.}]$ 
and a loop factor could 
modify it as $\frac{\lambda_5 }{8\pi^2}
\sum_j\frac{h_{\alpha j}h_{\beta j}}{M_{N_j}}f\left(\frac{M_{N_j}^2}{M_\eta^2}\right)$ 
where $M_\eta$ is the $\eta$ mass and 
$f(r_j)=\frac{r_j}{1-r_j}\left(1+\frac{r_j}{1-r_j}\ln r_j\right)$.\footnote{
In the present model, $\lambda_5$ is replaced by 
$\frac{\langle \sigma\rangle}{\Lambda}$ up to an $O(1)$ coupling constant.
The spontaneously generated $CP$ phase $\rho$ could induce 
a phase in the PMNS matrix \cite{l5}.}
Although they allow $M_{N_j}$ to take a value 
in TeV regions for the explanation of the
neutrino oscillation data, $M_{N_1}$ cannot be much reduced for successful 
leptogenesis in comparison with the type I seesaw case unless 
additional right-handed neutrino production processes exist \cite{ks1,ks2,lowlept1}. 
It is essential for low scale leptogenesis to introduce a new
production mechanism of the right-handed neutrinos in the thermal bath.
It is noticeable that this problem in low scale leptogenesis can be naturally resolved
 in the present model \cite{ps1,ps2,cpinf} since such an additional process is 
automatically built in the model. 

The vectorlike fermions could cause such a process through the mediation 
of $\sigma$ since the right-handed neutrino masses are also induced as
$M_{N_j}=y_{N_j}w$ through Yukawa couplings $y_{N_j}\sigma \bar N_jN_j^c$.
The vectorlike quarks $D_{L,R}$ can be in the thermal equilibrium 
as relativistic particles through gauge interactions at a temperature $T>M_D$. 
Then, the right-handed neutrinos can be generated effectively in the
thermal bath through the scattering 
$\bar D_LD_R\rightarrow \bar N_jN_j^c$, which is irrelevant to the neutrino Yukawa 
couplings $h_{\alpha j}$.
The reaction rate $\Gamma_{DD}^{(j)}$ of this scattering at the temperature 
$T$ is roughly estimated as
\begin{equation}
\Gamma_{DD}^{(j)}=\frac{y_D^2y_{N_j}^2}{4\pi}\frac{T^5}{\tilde\kappa_\sigma^2w^4}.
\end{equation}
If $H<\Gamma_{DD}^{(j)}$ is satisfied at a temperature $T$ which is
in a range $M_{N_j}<M_D<T< T_R$,
a sufficient amount of $N_j$ could be generated in the thermal bath. 
Thus, mother particles for the lepton number asymmetry could be prepared 
irrelevant to the largeness of $h_{\alpha j}$.

If we note that neutrino masses required by the neutrino oscillation data 
can be realized by two right-handed neutrinos, 
we find that the remaining neutrino Yukawa coupling $h_{\alpha 1}$ can take freely small 
values such as $O(10^{-8})$ independently from the constraints of the neutrino 
oscillation data.
In that case, the $N_1$ decay starts generating lepton number asymmetry 
at a low temperature for which washout processes of the generated lepton number 
asymmetry decouple. However, the $CP$ asymmetry $\varepsilon$ could be kept sufficient
magnitude.  
They make the sufficient production of the lepton number asymmetry 
possible even if $M_{N_1}$ is near TeV scales. 
The reheating temperature estimated in the previous part is found to be 
high enough for it. 

\begin{figure}[t]
\begin{center}
\begin{tabular}{ccccccc|c}\hline
$w$ & $u$ & $T_R$ & $y_{N_1}$ & $y_D$ & $\lambda_5$ & $h_1$ &   $Y_B(\equiv \frac{n_B}{s})$ 
\\ \hline
$10^6$& $10^8$ &$2\times 10^5$&
$1\times 10^{-2}$&$6.3\times 10^{-2}$&$3\times 10^{-5}$& $1\times 10^{-8}$ &
 $1.75\times 10^{-10}$  \\ \hline
$10^5$& $10^7$ &$2\times 10^4$&$1.5\times 10^{-2}$&$6.3\times 10^{-2}$&$1\times 10^{-5}$& $2\times 10^{-8}$ &
 $1.70\times 10^{-10}$ \\ \hline
\end{tabular}
\end{center}
\footnotesize{{\bf Table 2}~~Baryon number asymmetry predicted for typical parameters 
which give the explanation of the neutrino oscillation data. 
$y_{N_2}$ and $y_{N_3}$ are fixed as $y_{N_2}=2y_{N_1}$ and  $y_{N_3}=3y_{N_1}$.
$T_R$ corresponds to the lower bound value of the reheating temperature
expected for the assumed parameters.  A GeV unit is used as a mass scale.  }
\end{figure}

In Table 2, we give two numerical examples 
for the low scale leptogenesis with right-handed neutrino masses in the TeV range.
They are found by solving the Boltzmann equations in this model. 
In this calculation, we assume tribimaximal mixing for the 
PMNS matrix which can be realized by assuming that 
neutrino Yukawa coupling constants $h_{\alpha j}$ are represented 
by $h_j$ \cite{tribi}, respectively. 
$h_2$ and $h_3$ are fixed as values of $O(10^{-3})$ by using the neutrino 
oscillation data. 
In that case, the $CP$ asymmetry in the $N_1$ decay is found to be 
$\varepsilon\simeq -2\times 10^{-7}$ and the generated baryon number 
asymmetry is almost equal to the one which is caused from the lepton 
number asymmetry realized as $\varepsilon Y_{N_1}^{\rm eq}$ at 
a temperature $T(>M_{N_1})$. 
 As mentioned above,
this means that the washout processes play no substantial 
role since they almost decouple at the temperature 
where the $N_1$ decay starts. 
 
\section{Summary}
We have studied the inflation induced by an inflaton which is fixed by 
a $CP$ violating valley and the reheating caused after the end of inflation. 
The model is given as a simple extension of the SM with two singlet scalars 
and vectorlike quarks. If the $CP$ symmetry is spontaneously broken 
at a scale larger than the weak scale in the sector of the singlet scalars 
which are assumed to have the nonminimal coupling with the Ricci scalar, 
inflation is induced through the $CP$ violating valley. 
After the end of inflation, oscillating scalars produces
the excitation of the Higgs scalar through preheating. Since the produced Higgs 
excitations decay immediately so that they are not accumulated, 
nonlinear effects in the preheating do not play a role. As a result,
the reheating temperature cannot 
be high enough compared with $10^9$~GeV which is recognized 
as a lower bound for the successful thermal leptogenesis in the seesaw
frameworks.    

At the low energy regions, the singlet scalars relevant to the inflation 
could give an 
explanation for the origin of the $CP$ phase in the CKM matrix
through the mixing between the vectorlike quarks and the ordinary quarks.
This $CP$ phase does not contribute to the $\bar\theta$ parameter 
in the QCD sector since the mixing occurs in a way that the determinant of quark 
mass matrix is real. Thus, if the model is defined to have no explicit 
$CP$ violation, the strong $CP$ problem does not appear 
at tree-level at least. 
If vectorlike charged leptons are introduced to the lepton sector in the model,
a Dirac $CP$ phase in the PMNS matrix could be also explained in the same way. 
Since the scattering of the vectorlike fermions could effectively 
generate the right-handed neutrinos independently of the neutrino Yukawa 
couplings, the lightest right-handed neutrino could generate the sufficient 
lepton number asymmetry through the $CP$ asymmetric decay 
in a consistent way with the predicted low reheating temperature.
If the VEVs $w$ and $u$ takes values near TeV regions, the model might be 
examined through the unitarity violation of the CKM matrix and 
some FCNC processes in future precision experiments. 

\section*{Appendix A}
We briefly describe other processes producing 
$\chi,~\chi_\perp,\sigma$ at the zero crossing of $\chi$, where
$\chi_\perp$ corresponds to an angular component of $S$ in the case (i) 
and $S_R$ in the case (ii). 
As found from eq.~(\ref{imass2}), their couplings with $\chi$ satisfy 
$g_\chi/\hat\kappa_S=3,~ g_{\chi_\perp}/\hat\kappa_S= 1,~ 
g_\sigma/\hat\kappa_S \gg 1$, respectively.
Since these particles produced at the $\chi$ zero crossing are accumulated due to
the kinematical reason as discussed in the text, $\bar n_{\bar\Psi_k}$ shows an 
exponential behavior $e^{2\mu_k\tau}$.
However, they cannot play a crucial role because of the following reason.
In the case of $\chi$ and $\chi_\perp$, their excitations are produced fast 
but the process stops as soon as $\langle |\chi|^2\rangle$ 
and $\langle |\chi_\perp|^2\rangle$ reach a certain value such as $0.5\chi_{\rm end}^2/a^2$
\cite{smash, rescat}. Since the backreaction of these excitations to the inflaton oscillation 
restructures the resonance band,  the resonant particle production stops 
before causing much more conversion of the $\chi$ oscillation energy 
to the particle excitations. Moreover, since the decay of excitations produced through 
these processes are also closed kinematically, these could not play an efficient role in reheating. 
In the case of $\sigma$, the same reason as  $\chi$ and $\chi_\perp$ stops
the resonant production of its excitation at a certain stage
since $\sigma$ also couples with $\chi$ directly. 
These suggest that the main contribution to the preheating is expected 
to come from the $\phi$ producing process.

\begin{figure}[t]
\begin{center}
\footnotesize
\begin{tabular}{cccc|cccccccc}\hline
 &$u$ & $w$ & $M_D$ &$m_d$ & $m_s$ & $m_b$ & $|V_{ud}|$ & $|V_{ub}|$ & $|V_{sb}|$ & $J$ 
\\ \hline
(a) &$2\cdot 10^6$& $10^4$ &$1.9\cdot 10^3$&$3.0\cdot 10^{-3}$&
$8.8\cdot 10^{-2}$&
 $4.2$ & 0.21 & 0.0030 & 0.057 & $2.1\cdot 10^{-5}$ \\ \hline
(b) &$2\cdot 10^6$& $10^3$ & $1.9\cdot 10^3$ &$9.1\cdot 10^{-4}$&$6.1\cdot 10^{-2}$&
 4.2 & 0.22 & 0.0046 & 0.053 & $5.2\cdot 10^{-5}$    \\ \hline
(c) &$10^7$& $10^5$ & $1.1\cdot 10^4$ &$5.2\cdot 10^{-3}$&$8.2\cdot 10^{-2}$&
 4.3 & 0.21 & 0.0031 & 0.069 & $1.6\cdot 10^{-5}$    \\ \hline
\end{tabular}
\end{center}
\footnotesize{{\bf Table 3}~~Examples of the predicted values of 
mass eigenvalues of the down-sector quarks and elements of the CKM matrix. 
$J$ stands for the Jarlskog invariant \cite{j}. 
A GeV unit is used as a mass scale. 
Parameters in ${\cal M}_d$ are fixed at
$c=0.029$, $\lambda=0.25$, $p=0.18$, $q=0.038$, $r=0.5$ in (a), 
$c=0.03$, $\lambda=0.24$, $p=0.32$, $q=0$, $r=0.54$ in (b), 
and $c=0.03$, $\lambda=0.26$, $p=0.1$, $q=0.001$, $r=0.17$ in (c),  
respectively. The $CP$ violation tends to be larger for
 a larger $|{\cal F}_d|/M_D$ as expected. }
\end{figure}
\normalsize

\section*{Appendix B}
In this Appendix, we give concrete examples for the CKM matrix derived 
in this scenario in order to show that it can realize a substantial $CP$ phase.
We could find qualitative features of the scenario for the $CP$ phase 
generation discussed in the text through these examples.
A charged lepton mass matrix ${\cal M}_d$ in eq.~(\ref{ddmass}) is assumed to be 
\begin{equation}
{\cal M}_d=c\left(\begin{array}{ccc}
\lambda^4 & \lambda^3 & p\lambda^2 \\
\lambda^3 & \lambda^2 & q\lambda^2 \\
\lambda^2 & 1 & r \\
\end{array}\right),
\end{equation}
which has four real free parameters\footnote{This mass matrix has been studied in \cite{lmatrix} in other context before.}, and the diagonalization matrix 
$O_u$ for the up-sector 
is supposed to be a unit matrix in this study, for simplicity.
Yukawa couplings $y_{d_j}$, $\tilde y_{d_j}$ and $y_D$ in the sector of vectorlike fermions are assumed to be $y_d=(0, 6.3\times 10^{-4}, 0)$, 
$\tilde y_d=(0,0,-6.3\times 10^{-4})$ and $y_D=6.3\times 10^{-2}$, respectively.
The eigenvalues of the quark mass and the elements of the CKM matrix 
 are presented for assumed parameter sets in Table 3.
They suggest that the model could explain the origin of the $CP$ phase 
in the CKM matrix through the present inflation scenario.

\section*{Acknowledgements}
This work is partially supported by a Grant-in-Aid for Scientific Research (C) 
from Japan Society for Promotion of Science (Grant No. 18K03644).

\newpage
\bibliographystyle{unsrt}

\end{document}